\documentclass[12pt,preprint]{aastex}

\def\bnabla{\mbox{\boldmath{$\nabla$}}}
\def\bcross{\mbox{\boldmath{$\times$}}}

\title{Magnetic Rotator Winds and Keplerian Disks of Hot Stars}
\author{M. Maheswaran}
\affil{Department of Mathematics, University of Wisconsin-Marathon
County, \\ 518 S. 7th Avenue, Wausau, WI 54401}
\email{mmaheswa@uwc.edu}

\shorttitle{Magnetic Rotator Winds and Keplerian Disks}
\shortauthors{Maheswaran}

\begin{document}

\begin{abstract}
We set up equations and boundary conditions for magnetic rotator
winds and disks in axially symmetric hot stars in a steady state.
We establish a theorem stating that if a disk region has no
meridional motion but its magnetic field has a normal component at
a point $Q$ on its shock boundary, the angular velocity of the
disk region at $Q$ is the same as the angular velocity of the star
at the point $P_\star$ at which the magnetic field line through
$Q$ is anchored. When there is equatorial symmetry, all points of
the disk along the field line through $Q$ will have the same
angular velocity as $P_\star$. Also, we show that for a given
value of the magnetic field strength, if the rotation rate is too
high or the flow velocity into the shock boundary is too low, a
Keplerian disk region will not be formed. We consider the
formation of disks in magnetic rotators through the processes of
fill-up and diffusion into Keplerian orbits. At the end of the
fill-up stage the density of the disk increases significantly and
the magnetic force in the disk becomes negligible. If the
meridional component $B_m$ of the field at the surface is larger
than a minimum value $B_{m, min}$, a Keplerian disk can form. The
radial extent of the Keplerian region is larger for larger values
of $B_m$ and is largest when $B_m$ equals an optimal value $B_{m,
opt}$. The extent does not increase when $B_m$ is larger than
$B_{m, opt}$. If $\alpha$ is the ratio of rotational speed to the
critical rotation speed at the photosphere, the inner and outer
radii of the maximal quasi-steady Keplerian disk region are given
by $\alpha^{-2/3}R$ and $2^{4/3}\alpha^{-2/3}R$, respectively,
where $R$ is the stellar radius. For models with dipole-type
fields, the values of $B_{m, min}$ in B-type stars are of order 1
G to 10 G and in O-type stars they are about 500 G. Because the
values of $B_m$ required for disk formation in B-type stars are
relatively small and the fill-up time is short, we suggest that
meridional circulation may play a role in some of the
time-variation observed in disks of Be stars through its effect on
the magnetic field near the photosphere.

\end{abstract}

%\keywords{stars: winds, disks, magnetic fields, rotation}

\section{Introduction}
This paper has three main objectives. The first is to develop a
framework of equations and boundary conditions with general
validity for magnetic rotator winds and disks in hot stars. The
second is to determine the conditions under which Keplerian disk
regions may be formed. The third is to develop models to explain
the formation and properties of Keplerian disk regions. In a
future paper we intend to consider applications to Wind Compressed
Disk(WCD) regions. Earlier studies of magnetic rotator winds, such
as the one due to \cite{web67}, focussed on one-dimensional models
involving equatorial winds mainly to consider magnetic braking.
\citet{mes68} set up a system of equations for axially symmetric
stars with dipole fields. His equations were based on the
solutions obtained by \citet{cha56} for steady state
magnetohydrodynamic (MHD) systems. \citet{sak85} developed a
numerical method for two-dimensional magnetic rotator winds using
the Weber-Davis approach. There have been several other studies of
equatorial and non-equatorial winds and flows in the context of
disk formation, and a review of such work is given in a recent
paper by \citet{cas02} in which a Magnetically Torqued Disk (MTD)
model was proposed to explain the Be star phenomenon. An important
difference between the MTD model and our models is that we
consider weaker magnetic fields that are sufficient to provide the
torque required to sustain higher rotation rates in the flows
towards the disk region, whereas the fields discussed in the MTD
paper are strong fields required to spin up the disk. Since the
disk material is much denser than the the pre-shock flow material,
the fields required in the MTD model are stronger than the fields
required in our models.

Over the past decade, several models have been developed to
explain the formation of disks and equatorial density enhancements
around rapidly rotating stars. The idea of Wind Compressed Disks
in nonmagnetic stars was developed by \citet{bjo93} and the
concept of Wind Compressed Zones (WCZ) was presented by
\citet{ign96}. In the WCD model, trajectories of supersonic flow
material from a rotating star travel to the equatorial plane.
After the matter enters the disk region to form a WCD it continues
to flow outwards. However, the angular speed of this material is
far below the Keplerian orbital speed. An important similarity
between the WCD model and the magnetically channeled model is that
the flow passes through a compression shock which leads to a
significant enhancement of the density in the equatorial region.
The WCZ model is a milder version of the WCD model and does not
involve a magnetic field. Because it applies to stars that are
rotating too slowly to form WCD structures, the WCZ model does not
have a shock-compressed disk region and the enhancement of density
is less than in the WCD model. The WCZ idea has been modified by
\citet{ign98} to include the effects of a weak magnetic field. It
is called the WCFields model. In this model, the magnetic fields
are unimportant in accelerating the flow but an enhancement of the
field near the equatorial plane occurs through compression of
field lines. The WCFields model may be relevant in discussing the
flow in the equatorial region beyond and above the disk in
magnetic rotator models.

\citet{bab97a} developed a magnetically confined wind-shock (MCWS)
model to explain properties of Ap and Bp stars. In this model, a
strong dipole magnetic field channels the flow of wind from an
early type star to form a disk. \citet{bab97b} and \citet{don01}
have applied the MCWS model to hot stars in which observational
data have been interpreted to indicate the presence of dipole
magnetic fields with strengths in the range from 300 to 400 G and
with significant obliquity. Unfortunately, these papers neglected
the contribution of rapid rotation to the dynamics of the wind.
The hot stars of interest to us, especially Be-type, are known to
be extremely fast rotators and the effect of rotation on the
dynamics of the wind will be more significant than that of the
magnetic field. The MTD model addressed the question of disks in
fast rotators with strong magnetic fields.

Several phases in the development of disks around Be stars have
been revealed by long term observational studies of these objects.
Disks can form and disappear on time scales of about 30 years.
When the disks of Be stars achieve a certain mass, as estimated by
the equivalent width of the $\textrm{H}_\alpha$ emission line, the
stars tend to develop the observational phenomenon of $V/R$
variations. During the $V/R$ variation phase the star displays a
modification in which the $V$ component periodically changes so as
to appear stronger and weaker relative to the $R$ component. The
$V/R$ phenomenon has a typical time scale of about 7 years, which
has been interpreted by \citet{oka97} and others as the orbital
period of a one-armed spiral pattern and is typically much longer
than the orbital time of matter in the disk region around the
star.

In Section \ref{wind-zones}, we discuss the spatial features of
flow regions or wind zones in our models. In Sections
\ref{equations} and \ref{alfven-azimuthal} we set up systems of
equations that can be used to study properties of flows and winds
in rotating stellar models with magnetic fields. the models have
axial symmetry and are in a steady state. In a general study of
magnetic rotator winds it would be necessary to consider details
of the magnetic field over the entire stellar surface. If the
objective is restricted to studying the formation of disks, we
need to consider magnetic fields that are sufficiently strong only
over a smaller localized sector of the stellar surface through
which flux tubes emerge from the interior.

In Section \ref{jump-conditions} we set up jump conditions that
must be satisfied across the shock boundary between a flow region
and a disk region. We establish an important theorem in Section
\ref{rotn-theorem} connecting the angular velocity of a disk
region in which there is no meridional motion with the angular
velocity at the photosphere. In Section
\ref{condn-for-pre-kepl-disk} we consider conditions that must be
satisfied during the fill-up stage in a disk region with no
meridional motion and we derive a necessary condition for the
formation of a Keplerian disk. In Section \ref{disk-eqns} we set
up equations that govern the structure of a disk region in a
steady state. In Section \ref{rotation-of-disk-region} we discuss
the rotation of disk regions during the fill-up stage and, also,
the extent of Keplerian regions after a quasi-steady state has
been reached.

We consider two processes in the formation of Keplerian disk
regions. These are a fill-up process and a process of adjustment
of disk material with super-Keplerian rotation speeds into
Keplerian orbits. We use the term quasi-steady Keplerian disk to
describe a disk region that is formed after the processes of
fill-up and adjustment into Keplerian orbits. We study models in
which it is reasonable to consider the processes separately and we
discuss this in Sections \ref{rotation-of-disk-region},
\ref{results} and \ref{discussion}. Several authors
\citep{lee91,por99,oka01} have suggested that viscous stresses
play an important role in the structure and properties of disks.
We assume that viscous forces are not important during the fill-up
stage.

In Section \ref{mag-fld-constraints} we discuss some constraints
on the magnetic field and in Section \ref{mag-fld-models} we
develop a prescription for a stream function and the corresponding
dipole-type magnetic field configurations that enable us to study
disk formation. We discuss further details of our models and
develop a method of solution in Section \ref{models}. We present
our results in Section \ref{results}. Conclusions and possible
interpretation of observations are discussed in Section
\ref{discussion}.

\section{Flow Regions and Wind Zones}\label{wind-zones}
\begin{figure}
\epsscale {}
\plotone{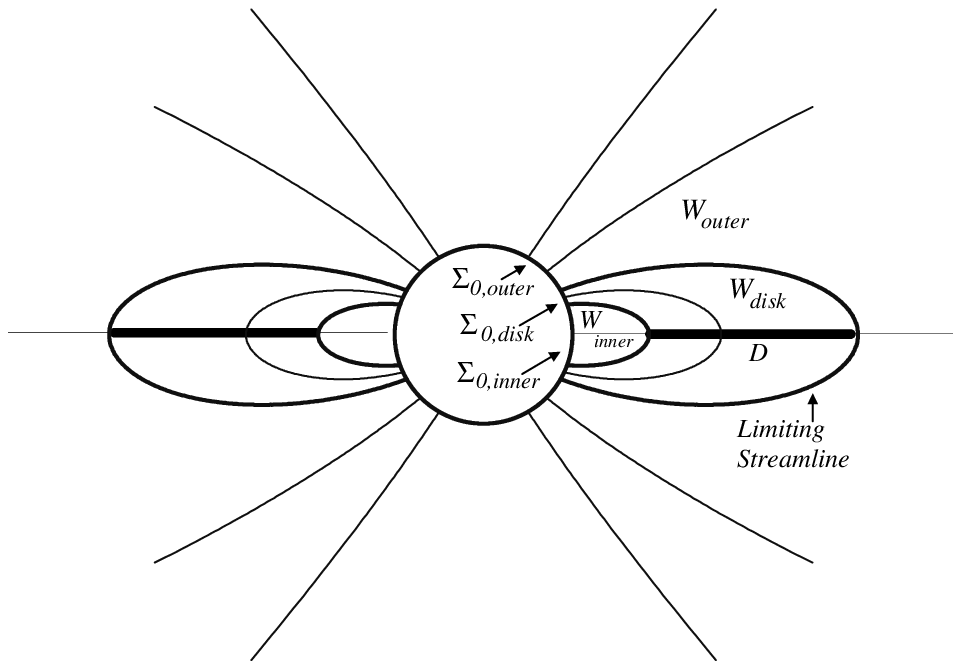} \caption{Schematic figure of
magnetic field lines, streamlines and equatorial disk during the
fill-up stage. Material from the sector $\Sigma_{0, disk}$ of the
initial surface travels through the wind zone $W_{disk}$  and
enters the disk region $D$. Material from $\Sigma_{0, outer}$
flows through the wind zone $W_{outer}$ out into space. Material
from the inner sector $\Sigma_{0, inner}$ flows into the wind zone
$W_{inner}$.\label{fig1}}
\end{figure}

We consider the flow of matter away from the stellar surface
caused by traditional wind driving processes. While most of the
flow from higher latitudes will travel to large distances away
from the star in the form of a wind, material from lower latitudes
can flow into equatorial disk regions or even flow back onto the
star. In order to simplify terminology, we to refer to any flow
region surrounding the star other than a disk region as a
{\textit{wind zone}}. This includes a region in which matter flows
to a disk region or back towards the stellar surface. Figure
\ref{fig1} shows a schematic drawing of a meridional cross-section
of the different wind zones and, also, the disk region during the
fill-up stage. In each hemisphere of our models, we consider three
wind zones. $W_{disk}$ denotes the wind zone in which the matter
flows into the disk region. $W_{outer}$ is the wind zone in which
material actually leaves the stellar environment in the form of a
wind. A third zone that may be present is $W_{inner}$ where matter
flows towards the equatorial plane and then back towards the
stellar surface. When disks are present, we consider models with a
single contiguous disk region $D$.

\placefigure{fig1}

We denote the photosphere by $\Sigma_\star$. Let $\Sigma_0$, which
we call the \textit{initial surface}, be a surface close to the
the photosphere on which initial conditions for the flow in the
wind zones are specified. We can take $\Sigma_0$ to be the sonic
surface or some other appropriate surface. Quantities associated
with the photosphere and the initial surface are denoted by
subscripts $\star$ and $0$, respectively. We identify different
sectors of the photosphere and the initial surface that are linked
to the different wind zones. At higher latitudes, we have the the
sector $\Sigma_{0, outer}$ of the initial surface from which
material flows through the wind zone $W_{outer}$ along open
streamlines to a large distance from the star. The meridional
speed of the flow in this zone reaches the meridional Alfv\'en
velocity at some distance from the star and exceeds this value at
points beyond the Alfv\'en distance so that the magnetic field
lines are drawn out by the flow. The sector of the photosphere
that is associated with $\Sigma_{0, outer}$ is denoted by
$\Sigma_{\star, outer}$. The flow from the sector $\Sigma_{0,
disk}$ of the initial surface travels through the wind zone
$W_{disk}$ and passes through a shock surface $\Sigma_{shock}$ to
supply material to the disk region $D$. Let $\Sigma_{\star, disk}$
be the sector of the photosphere that is linked to the sector
$\Sigma_{0, disk}$ by the magnetic field. If a third wind zone
$W_{inner}$ exists such that initially the flow is towards the
equatorial plane and then back towards the surface, we take
$\Sigma_{0, inner}$ and $\Sigma_{\star, inner}$ to be the sectors
of the initial surface and photosphere that are associated with
$W_{inner}$. If the sector $\Sigma_{\star, inner}$ is present, it
would be the closest to the equator. This zone does not play any
significant role in our models. The meridional speed of the flow
at all points in the regions $W_{inner}$ and $W_{disk}$ is less
than the meridional Alfv\'en speed so that the magnetic field is
able to influence the flow in theses zones. Although we separate
the flow into different zones, the basic equations that we set up
in Sections \ref{equations} and \ref{alfven-azimuthal} of this
paper are valid for all flow regions or wind zones.

An important feature in our models is the meridional streamline
between the wind zones $W_{outer}$ and $W_{disk}$. We use the term
{\it limiting streamline} to refer to this streamline, which
separates the flow entering the disk region from the wind flow to
large distances from the star. During the fill-up stage, the shock
boundary, $\Sigma_{shock}$, between the zone $W_{disk}$ and the
disk region $D$ does not extend outwards beyond the limiting
streamline. After the fill-up stage, the material in the disk
region with super-Keplerian rotation speed diffuses out until a
quasi-steady region is formed. The outer boundary of the
quasi-steady disk region may extend beyond the limiting streamline
and, if this happens, we assume that the wind material in the
region $W_{outer}$ flows outside the extended disk region.

We assume that, in addition to axial symmetry, streamlines of the
flow and magnetic field lines have equatorial symmetry. Therefore,
we need to set up our equations only in the \textit{Northern
Hemisphere}, which is determined by the direction of the angular
momentum vector. Equatorial symmetry together with appropriate
changes in sign, where necessary, will yield the values of
quantities and flow patterns in the Southern Hemisphere. In models
with surface magnetic fields that have predominantly dipole-type
structure we may assume, without loss of generality, that the
field and the flow in the Northern Hemisphere are in the same
direction and in opposite directions in the Southern Hemisphere.

\section{Equations for Flows and Winds}\label{equations}
In this section, we set up equations that are valid for the
material in all flow regions, or wind zones. We assume that the
flow is steady, which is reasonable when all other phenomena of
interest in the wind zones have time scales that are longer than
the characteristic time of the flow. We consider a rotating hot
star with a magnetic field. The star has mass $M$, luminosity $L$,
equatorial radius $R$ and effective temperature $T_{eff}$. For
simplicity, we assume that the star is approximately spherical and
that the system is axially and equatorially symmetric. We use a
stationary, inertial frame of reference with spherical polar
coordinates $(r, \theta, \phi)$. Let ${\mathbf{e}}_r$,
${\mathbf{e}}_\theta$, and ${\mathbf{ e}}_\phi$ be unit vectors in
the $r$-, $\theta$- and $\phi$- directions. Subscripts $m$ and
$\phi$ denote meridional and azimuthal components of vectors. The
velocity and magnetic field can be written in the form $\mathbf{v}
= \mathbf{v}_m + v_\phi \mathbf{e}_\phi$ and $\mathbf{B} =
\mathbf{B}_m + B_\phi \mathbf{e}_\phi$, respectively. Let $v_m$
and $B_m$ denote the magnitudes of $\mathbf{v_m}$ and
$\mathbf{B_m}$, respectively. The quantities $v_m$ and $B_m$ are
always positive because they are the magnitudes of vectors.
However, $v_\phi$ and $B_\phi$ can be positive or negative because
they are the $\phi$-components of the vectors.

\subsection{Kinematic Conditions for Steady Flow}
The density $\rho$ and the velocity $\mathbf{v}$ of the flow
material satisfy the continuity condition
\begin{equation}\label{div-rho-v}
{\rm div}(\rho{\mathbf v}) = 0 \, .
\end{equation}
The divergence-free condition for the magnetic field is
\begin{equation}\label{div-b}
{\rm div}{\mathbf B} = 0
\end{equation}
and the steady state MHD Induction Equation for material with
extremely high electrical conductivity is
\begin{equation}\label{curl-vxb}
{\rm curl}({\mathbf v}{\bcross}{\mathbf B}) = {\mathbf 0} \, .
\end{equation}
\cite{cha56} and \cite{mes68} have obtained useful integrals of
these differential equations. In an axially symmetric system, the
azimuthal component of Equation (\ref{curl-vxb}) requires that
\begin{equation}\label{vm-bm}
\mathbf{v}_m = \chi\mathbf{B}_m \, ,
\end{equation}
where $\chi = {(\mathbf{v}_m \, . \, \mathbf{B}_m)}/{B_m^2}$ is a
scalar function of position. Equation (\ref{vm-bm}) shows that the
meridional field lines coincide with the meridional streamlines.
Also, Equations (\ref{curl-vxb}) and (\ref{vm-bm}) give us
\begin{equation}\label{v-phi-1}
v_\phi = \chi B_\phi + \omega r\sin\theta \, ,
\end{equation}
where $\omega$ is a constant along a meridional streamline line.
We interpret $\omega$ to be the constant angular velocity of the
magnetic field line passing through the point $P_0$ at the initial
surface. Thus, if we picture the magnetic field line as being
anchored at the point $P_\star$ of the photosphere, then $\omega$
is the angular velocity of the star at $P_\star$. Equations
(\ref{vm-bm}) and (\ref{v-phi-1}) can be combined to give
\begin{equation}\label{v-total}
\mathbf{v} = \chi\mathbf{B} + \omega r \sin\theta \,
\mathbf{e}_\phi \, .
\end{equation}
Equations (\ref{div-rho-v}), (\ref{div-b}) and (\ref{vm-bm})
require that the quantity defined by  $\xi = \rho \chi$ must be
constant along a meridional streamline in a hemisphere. This
constant for a meridional streamline can be written
\begin{equation}\label{xi-eqn}
\xi = \rho\chi = \rho \, \frac{\mathbf{v_m} \, \cdot \,
\mathbf{B_m}}{B_m^2} \, .
\end{equation}
Note that $\chi > 0$ when $\mathbf{v}_m$ and $\mathbf{B}_m$ are in
the same direction and $\chi < 0$ when they are in opposite
directions. In Section \ref{azimuthal-wind-props} we discuss
properties of $v_\phi$ and $B_\phi$ and show that the quantity
$\chi B_\phi$ is negative in both hemispheres. Therefore, Equation
(\ref{v-phi-1}) implies that, relative to a stationary frame of
reference, the rotational velocity $v_\phi$ of a flow particle
moving along a field line is less than the co-rotation velocity
$\omega r \sin\theta$. If we consider points on a streamline
through the point $P_0(r_0,\theta_0,0)$ on the initial surface,
Equation (\ref{xi-eqn}) gives
\begin{equation}\label{vm-bm-1}
\frac{{\rho}v_m}{B_m} = \frac{\rho_0 \, v_{m,0}}{B_{m,0}} \, ,
\end{equation}
where subscript $0$ refers to values at $P_0$.

Equation (\ref{vm-bm}) has important consequences for modeling
stellar winds and disks. First, consider a star in which the
dominant component of the surface magnetic field is an odd
spherical harmonic, such as a dipole-type field. In both
hemispheres, material flows outwards from the stellar surface
while the the magnetic field is directed outwards in one
hemisphere and inwards in the other. Thus, $\chi$ must have
opposite signs in opposite hemispheres. This raises the question
of continuity of field lines across the equatorial plane. We
resolve this by noting that there are two possible scenarios for
the flow of material close to the equatorial plane. In one of
these, opposing flows from the different hemispheres will cross a
shock surface and form a disk in which the radial velocity is zero
or negligible. Here, shock conditions will account for the
discontinuity in $\chi$. In this paper, we consider models where
this condition prevails. In the other scenario with a dipole-type
field at the surface, the flow near the equatorial plane is
deflected parallel to the plane and we do not have to consider the
continuity of $\chi$ across the plane. If the flow crosses a shock
surface before moving parallel to the equatorial plane we expect a
WCD region to be formed. In both scenarios, the flow near the
equatorial plane in the region $W_{outer}$, outside the limiting
streamline, will be almost parallel to the plane so that
corresponding streamlines in opposite hemispheres are not
connected. Part of this flow may form a WCZ. In the case of a star
with a magnetic field in which the dominant component at the
initial surface is an even spherical harmonic, such as due to a
quadrupole field, the field lines near the equatorial plane in the
different hemispheres will be parallel to the plane and the
question of the continuity of $\chi$ does not arise.

\subsection{Dynamical Equations}
The general form of the steady state equation of motion for the
flow of matter is, in the usual notation,
\begin{equation}\label{momentum-eqn1}
\rho \, (\mathbf{v \cdot \bnabla})\mathbf{v} = -\bnabla p - \rho
\, \frac{(1-\Gamma) \, GM}{r^2}{\mathbf{e}}_r + \frac{{\rm curl}
\, \mathbf{B}\,{\bcross}\,\mathbf{B}}{4\pi} + \mathbf {F}_{line}
\, ,
\end{equation}
where $\Gamma = {\sigma_e L}/{(4\pi GMc)}$ is the ratio of the
stellar luminosity to the Eddington luminosity and $\sigma_e$ is
the electron scattering opacity. The term involving $\Gamma$ gives
the force due to continuum radiation written as a fraction of
local gravity. The force due to line radiation is $\mathbf
{F}_{line} = \rho \, g_L \mathbf{e}_r$, where $g_L$ is given by
the usual CAK expression \citep{cak75}. We assume that the
pressure is given by the isothermal equation of state
\begin{equation}\label{isothermal}
p = \rho \, c_s^2 \, ,
\end{equation}
which is a reasonable approximation for the wind zones and the
disk. Here, $c_s$ is the isothermal speed of sound. If we apply
the continuity condition (\ref{div-rho-v}) to the left hand side
of the $\phi$-component of Equation (\ref{momentum-eqn1}), and
then multiply both sides by $r \, \sin\theta$, we obtain
\begin{equation}\label{phi-eqn1}
{\rm div}(\rho \, \mathbf{v} \, r\sin\theta\,v_\phi \, ) =
{r\sin\theta} \, \left(\frac{{\rm curl} \, \mathbf{B}\bcross
\mathbf{B}}{4\pi} \right)_\phi \, .
\end{equation}
The right hand side of this equation represents the torque exerted
by the magnetic force on the flow material. This equation gives
the integral
\begin{equation}\label{phi-eqn2}
{r\sin\theta \, v_\phi} - \left(\frac{1}{4\pi\xi}\right)
r\sin\theta \, B_\phi = \mathcal{L} \, ,
\end{equation}
where $\mathcal{L}$ is a constant along a meridional streamline.
We interpret $\mathcal{L}$ to be the total angular momentum per
unit mass at points along the streamline.

In general, the line radiation force, $\mathbf{F}_{line}$, given
by the CAK expression is not conservative. In models where this
force is important, in order to obtain an energy equation we would
have to find a suitably defined potential function whose gradient
will be approximately equal to $\mathbf{F}_{line}$. Since our
primary interest is in understanding the effects of magnetic
fields, we consider models in which the line radiation force is
negligible and we set $\mathbf{F}_{line} = \mathbf{0}$. This is a
valid approximation in the case of late Be stars \citep{cas02}. We
can write Equation (\ref{momentum-eqn1}) in the form
\begin{equation} \label{momentum-eqn2}
\rho \, \bnabla  \left[\frac{1}{2} \, v^2 + c_s^2 \ln\rho -
\frac{(1-\Gamma) \, GM}{r} \right] = \rho \, \mathbf{v}\bcross
\left(\bnabla \bcross \mathbf{v}\right) + \frac{{\rm curl} \,
\mathbf{B} \bcross \mathbf{B}}{4\pi} \,.
\end{equation}
We take the scalar product of this equation with $\mathbf{v}$ and
use Equations (\ref{v-total}) and (\ref{phi-eqn1}) to obtain the
Bernoulli Equation
\begin{equation}\label{bernoulli}
\frac{v^2}{2} + c_s^2 \ln\rho - \frac{(1-\Gamma) \, GM}{r} -
\omega \, r\sin\theta \; v_\phi = {\mathcal{E}} \, ,
\end{equation}
where $\mathcal{E}$ is a constant along a meridional streamline
and has dimension of energy per unit mass.  The term $\, \omega \,
r\sin\theta \; v_\phi$ expresses the work done by the magnetic
torque on the flow material. If we move this term to the right
hand side of the equation and compare the new equation with the
corresponding equation for a nonmagnetic system, we see that the
energy of the flow material is increased by the work done by the
magnetic field. Only the azimuthal component of the magnetic force
does work on the material and supplies energy to it. This is
because the flow velocity, given by Equation (\ref{v-total}), is
the sum of a vector parallel to $\mathbf{B}$ and another in the
azimuthal direction while the magnetic force is perpendicular to
$\mathbf{B}$.

Our equations give us the set $\{\xi,\, \omega,\, \mathcal{L},\,
\mathcal{E}\}$ of streamline constants for the flow of material in
the wind zones. Here, $\xi$ is the ratio of the meridional mass
flux density to the meridional magnetic flux density and $\omega$
is the angular velocity of a flux tube associated with a
streamline. Also, $\mathcal{L}$ is the total angular momentum per
unit mass and $\mathcal{E}$ represents an energy per unit mass and
is not the total energy per unit mass of the flow material.

\section{Alfv\'en Velocities and Azimuthal Components}\label{alfven-azimuthal}
\subsection{Alfv\'en Velocities}\label{alfven-velocities}
Since the magnetic field plays an important role in determining
the nature of the flow, it is useful to compare the flow velocity
with the Alfv\'en velocity of the magnetic field, which is given
in vector form by ${\mathbf{B}}/{\sqrt{4\pi\rho}}$. We introduce
the quantities
\begin{equation}\label{alfven-mach-defn}
{\mathcal A}_m = \frac{B_m}{\sqrt{4\pi\rho} \; v_m} \qquad
\textrm{and} \qquad {\mathcal A}_\phi =
\frac{B_\phi}{\sqrt{4\pi\rho} \; v_\phi} \, .
\end{equation}
Note that ${\mathcal A}_m$ is always positive but ${\mathcal
A}_\phi$ can be positive or negative depending on the sign of
$B_\phi$. Also, ${\mathcal A}_m$ and $|{\mathcal A}_\phi|$ are,
respectively, the reciprocals of the meridional and azimuthal
Alfv\'en Mach numbers.

Streamlines flowing through the wind zone $W_{outer}$ are
associated with the material flowing away to large distances and
are characterized by the presence of an Alfv\'en point, $P_{alf}$,
at which $v_m$ equals the meridional component of the Alfv\'en
velocity. Let subscript $alf$ denote quantities at $P_{alf}$.
Then, we have
\begin{equation}\label{alfven-point}
v_{m,alf}^2 \, = \, \frac{B_{m,alf}^2}{4\pi\rho_{alf}} \, ,
\end{equation}
which satisfies the condition that ${\mathcal A}_{m,alf} = 1$.
Equations (\ref{vm-bm-1}) and (\ref{alfven-point}) combine to
yield
\begin{equation}\label{alfven-rho}
\rho_{alf} \, = \, \frac{4\pi\rho_0^2 \, v_{m,0}^2}{B_{m,0}^2} \,
= \frac{\rho_0}{{\mathcal A}_{m,0}^2}  \, ,
\end{equation}
which is the density at the Alfv\'en point on an open streamline
and is determined by the conditions at the initial surface.

In the wind zones $W_{disk}$ and $W_{inner}$, the magnetic field
is strong enough such the condition ${\mathcal A}_m
> 1$ is satisfied at all points in these zones. That is, the meridional
component of the flow velocity is sub-Alfv\'enic in these zones.
On the other hand, for a streamline in the wind zone $W_{outer}$
that emerges from the point $P_0$ on $\Sigma_{0, outer}$, we have
${\mathcal A}_m > 1$ at all points between $P_0$ and $P_{alf}$,
and ${\mathcal A}_m < 1$ for all points beyond $P_{alf}$. The
limiting streamline that separates the wind zones $W_{outer}$ and
$W_{disk}$ is the first meridional streamline above the equator
that contains an Alfv\'en point. Assuming that the system is
equatorially symmetric, the Alfv\'en point on the limiting
streamline will be on the equatorial plane.

For streamlines in all wind zones, we define two quantities
$\rho_c$ and $\ell$, which are constants for a given streamline
but can be different for different streamlines. The first is given
by the equation
\begin{equation}\label{critical-rho}
\rho_c \, = \frac{\rho_0}{{\mathcal A}_{m,0}^2} =
\frac{\rho}{{\mathcal A}_m^2}  \, .
\end{equation}
This may be interpreted as the characteristic density associated
with a streamline. The second quantity is defined by
\begin{equation}\label{critical-l1}
\omega \, \ell^2 = \mathcal{L} \, .
\end{equation}
We may interpret $\ell$ to be a characteristic distance from the
rotation axis associated with a given streamline. On an open
streamline in the region $W_{outer}$, we have $\rho_c =
\rho_{alf}$ and $\ell$ is the distance of $P_{alf}$ from the
rotation axis. For streamlines in the region $W_{disk}$ or
$W_{inner}$, we consider $\rho_c$ and $\ell$ to be streamline
parameters that enable us to conveniently express properties of
the flow. The value of $\ell$ for any streamline in the Northern
Hemisphere is given by Equation (\ref{critical-l2}) in Appendix
\ref{apndx-1}.

\subsection{Azimuthal Components}\label{azimuthal-wind-props} If
$\omega$ is the angular velocity at the point $P_\star$ of the
photosphere, we have $v_{\phi, \star} = \omega \, R \, sin
\theta_\star$. If $\theta_\star = \theta_0$, so that the points
$P_\star$ and $P_0$ are at the same latitude, Equation
(\ref{v-phi-1}) gives the relation between the rotational
velocities at the photosphere and the initial surface as
\begin{equation}\label{v-phi-star}
 \frac{v_{\phi,\star}}{R} = \frac{v_{\phi,0}}{r_0}\left(1 -
 \frac{B_{\phi,0}v_{m, 0}}{B_{m,0}v_{\phi,0}}\right)
 \, .
\end{equation}
Equations (\ref{v-phi-1}), (\ref{xi-eqn}), (\ref{phi-eqn2}) and
(\ref{critical-l1}) give us
\begin{equation}\label{v-phi-2}
v_\phi = \omega r \sin\theta \left[\frac{1-(\rho_c
/\rho)\{\ell^2/( r^2 \sin^2\theta )\}}{1-(\rho_c /\rho)} \right]
\end{equation}
and
\begin{equation}\label{b-phi-1}
B_\phi = 4\pi \rho \chi \, \omega r \sin\theta
\left[\frac{1-\{\ell^2/(r^2 \sin^2\theta )\}}{1-(\rho_c /\rho)}
\right] \, .
\end{equation}
Because $\rho > \rho_c$ when $r \, \sin\theta < \ell$ and vice
versa, Equation (\ref{b-phi-1}) requires that $B_\phi$ and $\chi$
must have opposite signs when $\omega > 0$. That is, $\chi B_\phi$
is always negative. In our models, $\chi$ is positive in the
Northern Hemisphere. Hence, $B_\phi$ and $\mathcal{A}_\phi$ are
negative in that hemisphere. Expressions for azimuthal components
and Alfv\'en velocities that are useful in applications to models
are given in Appendix \ref{apndx-1}.

The equations that we have set up are valid for any appropriate
rotation law at the stellar surface. We introduce the rotation
rate parameter $\alpha_\star = \alpha_\star(\theta_\star)$ for the
star at the point $P_\star$ at the stellar photosphere by taking
$\omega = \omega_{crit} \, \alpha_\star$, where
\begin{equation}\label{omega-max}
\omega_{crit}= \left[\frac{GM(1-\Gamma)}{R^3}\right]^{1/2} \, .
\end{equation}
Then, the rotational speed at the point $P_\star$ of the
photosphere is given by
\begin{equation}\label{v-phi-star-eqn}
v_{\phi, \star} = \omega_{crit} \, \alpha_\star \, R
\sin\theta_\star \, .
\end{equation}
From Equations (\ref{v-phi-star}) and (\ref{v-phi-star-eqn}) we
obtain
\begin{equation}\label{v-phi-0-eqn}
 v_{\phi, 0}= \omega_{crit} \, \alpha_\star \,
\left(\frac{\mathcal{A}_{m, 0}}{\mathcal{A}_{m, 0} -
\mathcal{A}_{\phi, 0}} \right)\, r_0 \, \sin\theta_0 \,
\end{equation}
for the rotational velocity of the flow material at the point
$P_0$ of the initial surface. In models where the angular velocity
is constant over the sector $\Sigma_{\star, disk}$ of the
photosphere, we write $\alpha_\star = \alpha$, where $\alpha$ is a
constant.

\section{Jump Conditions across the Shock Boundary}\label{jump-conditions}
The material from the sector $\Sigma_{0,disk}$ of the initial
surface passes through the shock surface $\Sigma_{shock}$ before
entering the disk region $D$. We consider jump conditions across
$\Sigma_{shock}$ during the fill-up stage of the disk. We assume
that the shock is isothermal, which is a reasonable approximation
for the types of models that we consider \citep{cas02}. In models
where the dominant component of the stellar magnetic field is an
odd spherical harmonic, such as a dipole-type field, we assume
that there is a region of the disk that has no meridional motion
during the fill-up stage. We refer to this pure rotational region
of the disk as a \textit{pre-Keplerian} region and we expect that
this region or part of it evolves into a quasi-steady Keplerian
disk region after the fill-up stage. On the other hand, in models
where the dominant component of the stellar field is an even
spherical harmonic, we expect that the material in the disk region
will have an outward flow along field lines that are parallel to
the equatorial plane. For convenience, we refer to such a region
as a WCD region. Another possible situation is one in which matter
in the inner part of the disk region is pre-Keplerian and the
outer part is of WCD-type. Although our main interest in this
paper is in pre-Keplerian regions, since one of our objectives is
to set up equations with general validity, we include conditions
for Pre-Keplerian and WCD regions.

In each hemisphere, $\Sigma_{shock}$ extends from the inner
boundary of the wind zone $W_{disk}$ to its outer boundary. In the
northern hemisphere of the meridional plane $\phi = 0$, the inner
boundary of $W_{disk}$ coincides with the streamline from the
point $P_{0, inn}$ on the initial surface to the point $Q_{inn}$
on $\Sigma_{shock}$. Let  $X_{inn}$ be the point on the equatorial
plane that is associated with $Q_{inn}$. The outer boundary of
$W_{disk}$ is the limiting streamline from $P_{0, lim}$ on the
initial surface to the point $Q_{lim}$ on $\Sigma_{shock}$ and is
associated with the point $X_{lim}$ on the equatorial plane. Let
$P_{0}(r_0, \theta_{0}, 0)$ be a general point on the sector
$\Sigma_{0, disk}$ so that $\theta_{0, lim} < \theta_0 <
\theta_{0, inn}$. The meridional streamline from $P_{0}$ travels
to the point $Q$ on the shock surface, which is associated with
the point $X(r_X, \pi/2, 0)$ on the equatorial plane. In models
where the disk region is a combination of a pre-Keplerian region
and a WCD region, the inner part of $D$ from $X_{inn}$ to the
point $X_{wcd}$ is pre-Keplerian and the outer part from $X_{wcd}$
to $X_{lim}$ is a WCD region. The point $X_{wcd}(r_{X, wcd},\,
\pi/2,\, 0)$ that separates the pre-Keplerian and WCD parts of $D$
is associated with the points $P_{0, {wcd}}(r_0, \theta_{0,
wcd},\, 0)$ on $\Sigma_{0, disk}$ and $Q_{wcd}$ on
$\Sigma_{shock}$. The equations that we set up in Section
\ref{equations} for the wind zones are also valid in a WCD region,
where there is a steady radial flow of material superimposed on
the rotational motion.

Let $\mathbf{n}$ be the unit normal vector to $\Sigma_{shock}$
from the wind zone into the disk. We define the unit tangent
vector $\mathbf{t}$ in the meridional plane as $\mathbf{t} =
\mathbf{n} \bcross \mathbf{e}_\phi$. We use subscript $n$ to
denote components of vectors in the direction of $\mathbf{n}$ and
the notation $\Delta[...]$ to represent the change in a quantity
across the shock boundary  We use a \textit{hat} to denote
quantities in the post-shock disk region and the subscript
\textit{fill} to indicate the values of quantities at the end of
the fill-up stage.

In a region of the disk with no meridional motion, the mass
increases steadily. In such a region, the density and the height
will change with time until a steady state is reached. The
magnetic field is assumed to be steady during the fill-up stage of
the disk. Let $h$ be the height of the shock surface at a radial
distance $r$ along the equatorial plane. Let $\delta$ be the angle
between the tangent line to the meridional cross-section of
$\Sigma_{shock}$ and the equatorial plane. In general, $\delta$
changes with $r$ and with time, and we assume that it is a small
positive angle. Then, $h$ increases with $r$ so that $\partial h
/\partial r = \tan \delta \geq 0$. Although $h$ and $\delta$
change with time, they do so on a time scale of the order of the
fill-up time $t_{fill}$ of the disk, which is much longer than the
flow time $t_{flow} = R / v_m$.

In both the pre-Keplerian and WCD regions of the disk,
conservation of magnetic flux requires that the normal component
of the magnetic field be continuous. Thus,
\begin{equation}\label{shock-bn}
\Delta \, [\, B_n \,] = 0 \, .
\end{equation}
Also, the tangential component of the electric field should be
continuous across $\Sigma_{shock}$. The electric field can be
written as $\mathbf{E} = -\mathbf{v} \bcross \mathbf{B}/c$.
Therefore, the jump condition requires that the tangential
component of $\mathbf{v} \bcross \mathbf{B}$ should be continuous
across the shock surface. The tangential component of $\mathbf{v}
\bcross \mathbf{B}$ has contributions from its $\phi$-component
and its $t$-component. The $\phi$-component of  $\mathbf{v}
\bcross \mathbf{B}$ is equal to $\mathbf{v}_m \bcross
\mathbf{B}_m$, which is equal to $\mathbf{0}$ in a flow region and
a WCD region because of Equation (\ref{vm-bm}), and in a
pre-Keplerian region because $\hat{\mathbf{v}}_m = \mathbf{0}$.
Therefore, for both pre-Keplerian and WCD regions the jump
condition reads
\begin{equation}\label{shock-et}
\Delta \, [ \, (\mathbf{v} \bcross \mathbf{B}) \cdot \mathbf{t}
\,] = 0 \, .
\end{equation}

The boundary conditions for mass flux are different for
pre-Keplerian and WCD regions. We consider them separately. In a
pre-Keplerian region, there is no radial motion and material
flowing into such a region will steadily increase the mass in that
region. This gives
\begin{equation}\label{shock-mass-k-1}
\rho \, v_n \, \sec\delta = \frac{\partial(h
  \hat{\rho})}{\partial t} \, .
\end{equation}
Because $\delta$ is quite small in thin disks and we can take
$\sec\delta \approx 1$. Also, as the disk region continues to fill
up, $\hat{\rho}_Q$ becomes much larger than $\rho_Q$ and the time
taken by $h$ to change significantly becomes quite long compared
with the time scale $t_{flow}$ for the flow of material towards
the disk.

The continuity of mass across the shock surface into a WCD region
gives
\begin{equation}\label{shock-vn-sk}
\Delta \, [\, \rho \, v_n \, ] = 0 \, .
\end{equation}
Then, for points $X$ between $X_{wcd}$ and $X_{lim}$ such that
$\theta_{0, lim} < \theta_0 < \theta_{0, wcd}$, since any inflow
from the pre-Keplerian part of the disk into the WCD part is
negligible over the time scales of interest to us, mass
conservation for flow into a WCD region requires that
\citep{bjo93}
\begin{equation}\label{shock-mass-sk-1}
  \int_{\theta_{0}}^{\theta_{0,wcd}}{2\pi r_0^2 \sin\theta \,
  \rho_0\,
  v_{r,0}} \, d\theta = 2\pi h \, r_X \; \hat{\rho} \,
\hat{v}_r \, ,
\end{equation}
where $\hat{v}_r$ is the speed of the WCD flow parallel to the
equatorial plane averaged over the height of the disk. In models
where $\rho_0$ and $v_{r,0}$ are constant, this condition
simplifies to
\begin{equation}\label{shock-mass-sk-2}
  r_0^2 \, \rho_0 \, v_{r,0}\left(\cos\theta_{0} -
  \cos\theta_{0,wcd}\right) = h \, r_X \; \hat{\rho}\, \hat{v}_r \, .
\end{equation}
Equations (\ref{shock-mass-sk-1}) and (\ref{shock-mass-sk-2}) give
the rate of mass flow in the Northern Hemisphere only. The total
rate of mass flow is twice the value of the integral.

If we substitute $\textrm{curl} \, \mathbf{B} \bcross \mathbf{B} =
(\mathbf{B} \cdot \bnabla) \mathbf{B} - \bnabla (B^2 )/2$ in
Equation (\ref{momentum-eqn1}), the continuity condition for
momentum flux across $\Sigma_{shock}$ into a pre-Keplerian or WCD
region yields
\begin{equation}\label{shock-mom}
\Delta \, \left[ \, \rho v_n \, \mathbf{v} + \left(p +
\frac{B^2}{8\pi}\right)\mathbf{n} - \frac{B_n \, \mathbf{B}}{4\pi}
\,\right] = \mathbf{0} \, ,
\end{equation}
where we have assumed that the line radiation force is negligible.

Next, we consider the conditions that must be satisfied by the
streamline constants $\xi$, $\omega$, $\mathcal{L}$ and
$\mathcal{E}$ for flow from the wind zone into a WCD region across
the surface $\Sigma_{shock}$. Equations (\ref{shock-bn}),
(\ref{shock-vn-sk}) and (\ref{xi-eqn}) imply that
\begin{equation}\label{shock-rho-xi}
\Delta \, \left[\, \xi \,\right] = 0.
\end{equation}
Also, Equations (\ref{v-total}), (\ref{shock-et}) and
(\ref{shock-bn}) together with the fact that $\mathbf{e}_\phi
\bcross \mathbf{t} = \mathbf{n}$ require that
\begin{equation}\label{shock-omega}
\Delta \, \left[\, \omega \, \right] = 0.
\end{equation}
Equation (\ref{phi-eqn2}) and the azimuthal component of
(\ref{shock-mom}) ensure that
\begin{equation}\label{shock-L}
\Delta \, \left[\, \mathcal{L} \, \right] = 0.
\end{equation}

In general, the streamline constant $\mathcal{E}$ in the Bernoulli
Equation is not continuous across the shock boundary of the disk
region. We do not consider jump conditions for streamline
constants across the shock boundary of a pre-Keplerian region
because there is no meridional motion there. However, since
$\omega$ is associated with the rotation of a magnetic field line,
in Section \ref{rotn-theorem} we discuss the continuity of
$\omega$ across the shock boundary of a pre-Keplerian region.

\section{A Theorem on Disk Rotation}\label{rotn-theorem}
Here, we consider the jump condition given by Equation
(\ref{shock-et}) at a point $Q(r_Q, \theta_Q, 0)$ on the shock
boundary of a disk region with no meridional motion. In the
pre-shock the flow region, Equation (\ref{v-total}) gives
$(\mathbf{v} \bcross \mathbf{B}) \cdot \mathbf{t} = - \omega \,
r_Q \sin\theta_Q \, B_{n, Q}$ . In the disk region we have
$\hat{\mathbf{v}} = \hat{v}_\phi \mathbf{e}_\phi$ and this gives
$(\hat{\mathbf{v}} \bcross \hat{\mathbf{B})} \cdot \mathbf{t} = -
\hat{v}_{\phi, Q} \, \hat{B}_{n, Q}$. Also, Equation
(\ref{shock-bn}) states that $B_{n, Q} = \hat{B}_{n, Q}$.
Therefore, Equation (\ref{shock-et}) gives
\begin{equation}\label{shock-vphi-jump}
\hat{v}_{\phi, Q} = \omega \, r_Q \sin\theta_Q \, .
\end{equation}
This equation shows that the point in the disk region that is
adjacent to $Q$ rotates with angular velocity $\omega$. In Section
\ref{equations} we noted that the angular velocity $\omega$
associated with a field line is the same as that of the point
$P_\star$ of the star where the field line anchored. Thus, we have
an extremely important result for the rotation in disk regions
where the meridional speed is negligible. We state this in the
form of a theorem as follows:
\newtheorem{theorem}{Theorem}
\begin{theorem}\label{thoerem-disk-rotn}
Let $Q$ be any point on the shock boundary between a magnetic
rotator flow region and a disk region of an axially symmetric star
that is in a steady state. Suppose that there is no meridional
motion in the disk region and that the normal component of the
magnetic field in the disk region at $Q$ is not zero. Let
$\hat{Q}$ be the point of the disk region adjacent to $Q$ on the
field line through $Q$. Then, the angular velocity of the disk
region at $\hat{Q}$ is equal to the angular velocity, $\omega$, at
the point $P_\star$ of the star at which the magnetic field line
through $Q$ is anchored. \newline Furthermore, if the system is
equatorially symmetric and the magnetic field line through $Q$ is
continuous across the disk, then the angular velocity at all
points of the disk region along the field line will be the same as
that at $P_\star$ for time periods over which the disk region
remains steady.
\end{theorem}
The first part of the theorem follows from Equation
(\ref{shock-vphi-jump}) and requires that the meridional speed in
the disk region be zero. To establish the the second part we use
the equation for the pre-Keplerian region that corresponds to
Equation (\ref{v-phi-1}). Since $\hat{\mathbf{v}}_m = \mathbf{0}$,
we have $\hat{\chi} = 0$ and we obtain $\hat{v}_\phi =
\hat{\omega} r\sin\theta$, where $\hat{\omega}$ is constant along
a meridional field line. Comparing this equation for
$\hat{v}_\phi$ with Equation (\ref{shock-vphi-jump}) we see that
$\hat{\omega} = \omega$ for the field line through $Q$. Thus, all
points of the pre-Keplerian region along a field line will have
the same angular velocity as the point of the star where the field
line is anchored over time scales where the region is steady.
During the fill-up stage, the flow time, $t_{flow}$, is much
shorter than the the fill up time, $t_{fill}$, so that over time
scales of the order of $t_{flow}$ the second part will be true. A
situation in which the second part of the theorem will be true
over longer periods of time that are comparable to $t_{fill}$ is
when the region $\Sigma_{\star, disk}$ is in uniform rotation.
Then the corresponding pre-Keplerian region of the disk will have
the same constant angular velocity as that sector of the star. The
models that we consider in our applications satisfy this
condition.

We emphasize that our theorem is quite different from Ferraro's
Law of Isorotation although part of it may appear to be similar.
We require the presence of a shock surface and in the region
between the stellar surface and the shock surface there is
meridional motion as well as differential differential rotation of
the outflow material. In models satisfying Ferraro's Law, we can
easliy show that there should be no meridional motion outside the
stellar surface. From our discussion in Section \ref{equations},
we know that $\chi$ must be constant along field lines and have
opposite signs in opposite hemispheres. If there is no shock
surface then the continuity of $\chi$ across the equatorial plane
requires that $\chi = 0$ along every field line. Therefore, $v_m$
must be zero along every field line and there can be no meridional
motion anywhere in an atmosphere where Ferraro's law is satisfied.

\section{Conditions for the Formation of Pre-Keplerian Disk Regions}
\label{condn-for-pre-kepl-disk} In applications to models, we
consider conical shock surfaces given by $\theta = (\pi/2) -
\delta$ in the Northern Hemisphere, where $\delta$ is small and
has the same value for all $r$. It may increase steadily with time
during the fill-up stage. In such a model, the cross-section of
$\Sigma_{shock}$ is a straight line given by $h = r \, \tan
\delta$ and we have $\mathbf{n} = \mathbf{e}_\theta$ and
$\mathbf{t} = \mathbf{e}_r$. The jump conditions that we set up in
Section \ref{jump-conditions} give us the following equations at
the point $Q$ on the shock boundary of a pre-Keplerian region of
the disk, where $\hat{v}_r = 0$ and $\hat{v}_\theta = 0$:
\begin{eqnarray}
\label{kepl-disk-bdry-v-phi} \hat{v}_{\phi, Q} & = & v_{\phi, Q} -
\left(\frac{v_{\theta, Q}}{B_{\theta, Q}} \right)B_{\phi, Q} \, =
\, \omega \, r_Q \, \sin\theta_Q \, .
\\
\label{kepl-disk-bdry-B-theta} \hat{B}_{\theta, Q} & = &
B_{\theta, Q} \, .
\\
\label{kepl-disk-bdry-B-r} \hat{B}_{r, Q} & = & B_{r, Q} - 4\pi
\xi \, v_{r, Q} \, = \, B_{r, Q}\left(1 - \frac{1}{\mathcal{A}_{m,
Q}^2} \right) \, .
\\
\label{kepl-disk-bdry-B-phi} \hat{B}_{\phi, Q} & = & B_{\phi, Q} -
4\pi \xi \, v_{\phi, Q} \, = \, B_{\phi, Q}\left(1 -
\frac{1}{\mathcal{A}_{m, Q} \, \mathcal{A}_{\phi, Q}} \right) \, .
\end{eqnarray}
Equation (\ref{kepl-disk-bdry-B-phi}) can be combined with
Equation (\ref{phi-eqn2}) to obtain
\begin{equation}\label{disk-B-phi}
\hat{B}_{\phi, Q} = B_{\phi, 0}\left(\frac{r_0 \sin\theta_0}{r_Q
\sin\theta_Q}\right)\left(1 - \frac{1}{\mathcal{A}_{m,
0}\mathcal{A}_{\phi, 0}}\right) \, .
\end{equation}

At the end of the fill-up stage, if $\hat{\rho}_{fill}$ is the
disk density averaged over the height of the disk at any point, we
can use the jump condition (\ref{shock-mom}) for momentum flux to
write
\begin{equation}\label{disk-density-1}
\hat{\rho}_{fill, Q} = \left[\rho\left( \frac{c_s^2 +
v_\theta^2}{\hat{c}_s^2} \right) + \left(\frac{B^2 -
\hat{B}^2}{8\pi \hat{c}_s^2} \right)\right]_Q \, .
\end{equation}
Equation (\ref{kepl-disk-bdry-B-r}) shows that $\hat{B}_{r,Q}^2$
is slightly less than $B_{r,Q}^2$ and Equation
(\ref{kepl-disk-bdry-B-phi}) shows that $\hat{B}_{\phi,Q}^2$ is
slightly larger than $B_{\phi,Q}^2$. In stellar models with
dipole-type fields, we have $B_{r,Q} \approx 0$ near the
equatorial plane and because of Equation
(\ref{kepl-disk-bdry-B-theta}) we can write $B^2 - \hat{B}^2 =
B_{\phi,Q}^2 - \hat{B}_{\phi,Q}^2$, which will be negative. Hence,
in such models, the density of a Keplerian disk will be somewhat
less than that in a nonmagnetic model. Also, if the magnetic field
present is too strong, Keplerian disk formation may not be
possible. A necessary condition for disk formation can be obtained
by requiring that $\hat{\rho}_{fill} > 0$ in Equation
(\ref{disk-density-1}). Combining this with Equation
(\ref{kepl-disk-bdry-B-phi}), using the relation $B_{\phi}^2 = 4
\pi \rho \, v_{\phi}^2 \, \mathcal{A}_{\phi}^2$ and noting that
$v_\theta \gg c_S$ near the shock surface, we obtain a condition
for Keplerian disk formation as
\begin{equation}\label{disk-formation-condn}
\frac{v_{\theta, Q}}{v_{\phi, Q}} > \left(\frac{1 - 2 A_{m,
Q}A_{\phi, Q}}{2 \mathcal{A}_{m, Q}^2}\right)^{1/2} \, .
\end{equation}
That is, for a given rotational velocity of the pre-shock flow at
a point of the shock surface, the normal component of the flow
velocity at that point must be larger than a value determined by
the magnetic field. Also, if $v_{\theta, Q}$ and $v_{\phi, Q}$ are
directly related to the wind terminal velocity, $v_\infty$, and
the angular velocity, $\omega$, of the star , respectively, we
expect that for a given value of $v_\infty$ there will be an upper
bound on $\omega$ beyond which a Keplerian disk will not form.

We define the fill-up time at radial distance $r$ for a
pre-Keplerian disk region, to be
\begin{equation}\label{fill-up-time-1}
t_{fill}= \frac{h_{fill} \, \hat{\rho}_{fill}}
{{\partial(h\hat{\rho})}/{\partial t}} \, .
\end{equation}
Our expression for $t_{fill}$ is different from the one used by
\citet{cas02}. Since $\delta$ is small during the fill-up stage
for thin disks, we take $\cos\delta \approx 1$. Then, Equations
(\ref{shock-mass-k-1}) and (\ref{fill-up-time-1}) give us
$t_{fill} = {h_{fill} \, \hat{\rho}_{fill}}/{\rho v_n}$. For our
conical shock surface with $h_{fill} = r \tan\delta_{fill}$, we
have
\begin{equation}\label{fill-up-time-2}
t_{fill} = \left( \frac{\hat{\rho}_{fill}}{\rho }\right)
\,\left(\frac{r \, \tan\delta_{fill}}{v_\theta}\right)  \, ,
\end{equation}
where the value of $\hat{\rho}_{fill}$ is given by Equation
(\ref{disk-density-1}). After the fill-up stage, the density of a
Keplerian disk region will be sufficiently large and we assume
that the disk temperature will be determined essentially by
stellar radiative heating.

\section{Equations for Disk Regions}\label{disk-eqns}
In this section we consider equations for a disk region $D$ that
is in a quasi-steady state. We assume that $v_\theta = 0$ in $D$
and the equations for the disk region are written for values of
quantities averaged over the height of the disk. The $r$- and
$\theta$- components, respectively, of the momentum equation for
the disk are
\begin{equation}\label{disk-mom-eqn-r}
\hat v_r \frac{{\partial \hat v_r }}{{\partial r}} = - \,
\frac{1}{\hat{\rho}} \, \frac{{\partial \hat {p}}}{{\partial r}} +
f_r + \mathbf{F}_{line} \cdot {\mathbf{e}}_r \, .
\end{equation}
and
\begin{equation}\label{disk-mom-eqn-phi}
\frac{\hat v_r}{r} \, \frac{\partial}{\partial r}\left(r\hat
v_\phi\right) =  {\mathcal{F}}_{\phi} - \frac{1}{h \hat{\rho} \,
r^2} \frac{\partial} {\partial r}\left(r^2 \nu h \hat{\rho} \,
c_S^2\right)\, .
\end{equation}
Here,
\begin{equation}\label{disk-force-r}
f_r = \frac{\hat{v}_{\phi}^2}{r} - \frac{GM(1-\Gamma)}{r^2} +
{\mathcal{F}}_{r} \, ,
\end{equation}
\begin{equation}\label{disk-mag-force-r}
{\mathcal{F}}_{r} = \frac{1}{4\pi\hat{\rho} \, r}
\left[\hat{B}_\theta \frac{\partial}{\partial \theta}(\hat{B}_r) -
\hat{B}_\theta\frac{\partial}{\partial r}(r\hat{B}_\theta) -
\hat{B}_\phi\frac{\partial}{\partial r}(r\hat{B}_\phi) \right] \,
\end{equation}
and
\begin{equation}\label{disk-mag-force-phi}
{\mathcal{F}}_{\phi} = \frac{1}{4\pi\hat{\rho} \, r}
\left[\hat{B}_r \frac{\partial}{\partial r}(r \hat{B}_\phi) +
\frac{\hat{B}_\theta}{r \, \sin\theta} \, \frac{\partial}{\partial
\theta} \, (r \, \sin\theta \, \hat{B}_\phi) \right] \, .
\end{equation}
The last term in Equation (\ref{disk-mom-eqn-phi}) represents the
effect of viscosity (e.g., Okazaki 2001) and $\nu$ denotes the
viscosity coefficient, which has a value of order 0.1. The forces
per unit mass due to the magnetic field in the radial and
azimuthal directions are given by ${\mathcal{F}}_{r}$ and
${\mathcal{F}}_{\phi}$, respectively. The quantity $f_r$ is the
resultant of the three main forces (per unit mass) that control
the radial momentum of the disk. These equations are supplemented
by a mass conservation equation that has the form $4 \pi r h
\hat{\rho} \hat{v}_r = \dot{M}_{disk}$, which represents the
amount of mass flowing across a cross-section of the disk at
radius $r$. In WCD regions, this equation can be replaced by
Equation (\ref{shock-mass-sk-1}). In a strict Keplerian disk
$\hat{v}_r $ and $\dot{M}_{disk}$ are zero.

In a Keplerian region of $D$ we assume that $\hat{v}_r$ and
$\bnabla \hat{p}$ are negligible. In stars where the force due to
line radiation is not significant we take $\mathbf{F}_{line} =
\mathbf{0}$. Then, Equation (\ref{disk-mom-eqn-r}) requires that
$f_r = 0$ and Equation (\ref{disk-force-r}) gives us an expression
for the Keplerian rotation speed, $\hat{v}_{\phi, K}$, in the form
\begin{equation} \label{disk-Kepl-rotn-vel}
\hat{v}_{\phi, K} = \left[\frac{GM(1 - \Gamma)(1 -
\zeta)}{r}\right]^{1/2} \,
\end{equation}
where $\zeta = {r^2 {\mathcal{F}}_{r} }/{\{GM(1-\Gamma)\}}$ is the
ratio of the radial component of the magnetic force to {\it
effective gravity}. Note that $\zeta$ becomes negative when
$\mathcal{F}_{r}$ is directed inwards and $\zeta \approx 0$ when
the magnetic force in the disk is small compared to effective
gravity. Also, in a Keplerian disk region Equation
(\ref{disk-mom-eqn-phi}) requires that ${\mathcal{F}}_{\phi} -
\{{\partial} \left(r^2 \nu h \hat{\rho} \, c_S^2\right)/{\partial
r}\}/({h \hat{\rho} \, r^2}) = 0$. The results presented for our
models in Section \ref{results} show that the magnetic force and
the viscous force become negligible at the end of the fill-up
stage. Thus, both terms on the right hand side of Equation
(\ref{disk-mom-eqn-phi}) become negligible. Although we do not
study the evolution of the disk region after a quasi-steady
Keplerian region is established, if a viscous force becomes
significant, then in order to maintain Keplerian rotation, either
the magnetic torque on the disk material should offset any
decrease in angular momentum due to viscosity or there should be a
continuous supply of angular momentum to the disk material. In the
case of a WCD region, the steady state equations are essentially
the same as those that we derived in Sections \ref{equations} and
\ref{alfven-azimuthal} for the wind zones with the additional
conditions that $\hat{v}_\theta = 0$ and $\hat{B}_\theta = 0$.

\section{Rotation and Extent of Disk Regions}\label{rotation-of-disk-region}
We introduce the non-dimensional variable $x = {r_X}/{R}$ to
represent the radial distance of the general point $X(r_X,
{\pi}/{2}, 0)$ on the equatorial plane. We consider a
pre-Keplerian disk region with shock surface given by $\theta =
(\pi/2)-\delta$ in the Northern Hemisphere. Suppose that $Q(r_Q,
(\pi/2)-\delta, 0)$ is the point on the shock surface such that
the meridional field line through $Q$ passes through $X$. When the
disk is thin we assume that $Q$ is vertically above $X$. Then, we
have $r_X = r_Q \, \cos\delta$. This is a good approximation in
the case of a thin disk with a dipole-type field because the field
line through $X$ is perpendicular to the equatorial plane at that
point. We assume that all points in a narrow vertical column
through $X$ in the disk region have the same angular velocity.

\subsection{Rotation of a Pre-Keplerian Region During the Fill-up Stage}
\label{pre-kepl-exact-kepl} Theorem \ref{thoerem-disk-rotn}, which
we derive in Section \ref{rotn-theorem}, gives
\begin{equation}\label{disk-v-phi-soln-1}
\hat{v}_\phi = \omega \, r = \omega_{crit} \, \alpha_\star \, r
\end{equation}
for the rotational velocity of disk at distance $r$ from the
center of the star during the fill-up stage. The angular velocity,
$\omega$, is constant along a field line and can have different
values along different field lines. Thus, along the equatorial
plane, $\omega$ is a function of $r$. To obtain this function we
must know $\omega$ as a function of colatitude $\theta$ at the
photosphere and, also, the stream function, $\psi(r, \theta)$, for
the field lines linking the points $P_\star$ on the sector
$\Sigma_{\star, disk}$ to the points $X$ in the disk region $D$.
However, even when the stream function is not known, we are able
to discuss the rotation of a pre-Keplerian disk region of a star
where the sector $\Sigma_{\star, disk}$ has an approximately
constant angular velocity $\omega$. In this case, the
corresponding pre-Keplerian disk region will have the same
constant angular velocity. Before we consider models with simple
rotation laws at the photosphere, we consider the question of
whether it possible for exact Keplerian rotation to be established
in a pre-Keplerian region at the end of the fill-up stage. We look
at two examples which show that this is very unlikely in real
stars.

In a general model, the condition for Keplerian rotation of the
disk is obtained using Equations (\ref{disk-Kepl-rotn-vel}),
(\ref{disk-v-phi-soln-1}) and (\ref{omega-max}). It reads
\begin{equation}\label{disk-kepl-condn-1}
x^3 \, \alpha_\star^2  = 1 - \zeta \, ,
\end{equation}
where $\zeta$ is function of $r$ and $\alpha_\star$ is a function
of $\theta_\star$. Suppose that we consider an idealized model in
which the angular velocity distribution at points in the sector
$\Sigma_{\star, disk}$ of the photosphere results in a
pre-Keplerian disk region that has exact Keplerian rotation.
Although such a possibility may be mainly of theoretical interest,
it is quite instructive. Also, suppose that the disk is at the end
of the fill-up stage when $\zeta$ has become negligible. In such a
model, Equation (\ref{disk-kepl-condn-1}) requires that
$\alpha_\star = x^{-3/2}$ for exact Keplerian rotation at a point
$X$ of the disk. Suppose that all points of the disk from $X_{1}$
to $X_{2}$ are in exact Keplerian rotation and let $P_{\star, 1}$
and $P_{\star, 2}$, respectively, be the points at the photosphere
that are linked to $X_{1}$ and $X_{2}$ by the magnetic field. Let
us select $x_{1} = 1.3$ and $x_{2} = 3.5$. Then, the rotation rate
$\alpha_\star$ at points in the sector $\Sigma_{\star, disk}$ of
the photosphere must decrease from $0.67$ at $P_{\star, 1}$ and
$0.15$ at $P_{\star, 2}$. We do not expect that the rotation rate
at the photosphere in real stars will decrease so drastically for
a relatively small increase in latitude. Thus, for realistic
angular velocity distributions at the photosphere, we expect that
if the rotational velocity of the pre-Keplerian disk has a
Keplerian value at a point $X_{kep}$ then the rotation rate will
be super-Keplerian at points between $X_{kep}$ and $X_{lim}$.

Another hypothetical model with exact Keplerian rotation is one in
which the magnetic field in the disk region is quite strong and
has a configuration such that the magnetic force is able to offset
the difference between the centrifugal force and gravity. However,
for the types of models that are of interest to us, the magnetic
field required will be far too strong. This is because the radial
component of the magnetic force in the disk must satisfy the
condition $\zeta = 1 - x^3 \, \alpha^2$ and, for example, if a
star with $\alpha = 0.6$ is to have a pre-Keplerian disk extending
to a distance of $x = 3.5$, the equations that are given in
Section \ref{disk-eqns} together with the disk density at the end
of the fill-up stage given by Equation (\ref{disk-density-1}) show
that the value of $\zeta$ must be approximately equal to $-14$.
Such a large value of $\zeta$ in the disk region will require an
exceptionally strong magnetic field at the photosphere. Also, the
topology of the field would have to be highly contrived to produce
a magnetic force in the disk region that has the required
magnitude and direction.

The different cases that we have considered here suggest quite
strongly that the most plausible scenario for the formation of a
quasi-steady Keplerian disk region is one where a pre-Keplerian
disk region is formed with material having super-Keplerian
rotation speeds and then this material diffuses into Keplerian
orbits when the magnetic force becomes small.

\subsection{Radial Extent of Quasi-steady Keplerian Disks} \label{kepl-radial-extent}
Consider a model in which the sector $\Sigma_{\star, disk}$ of the
photosphere is in uniform rotation with angular velocity $\omega$.
Let the surface magnetic field be sufficiently strong for the
formation of a pre-Keplerian region of the disk during the fill-up
stage. If there is no meridional motion in the disk region, our
theorem implies that the angular velocity of this pre-Keplerian
disk region will be constant and equal to $\omega$. Suppose that
the surface field is not too strong and that at the end of the
fill-up stage the disk density $\hat{\rho}_{fill}$ becomes
sufficiently large such that the magnetic force in the disk
becomes negligible compared with the centrifugal force. Let
$X_{kep}$ be the innermost point of the disk where the rotation
speed is Keplerian. To simplify our discussion, we take $X_{kep}$
to be at the inner boundary of the pre-Keplerian region of the
disk. Let $x_{kep}$ be the value of $x$ at the point $X_{kep}$.

If $\alpha$ is the rotation rate of the sector $\Sigma_{\star,
disk}$, Equations (\ref{v-phi-star-eqn}) and
(\ref{disk-kepl-condn-1}) give
\begin{equation}\label{disk-kepl-X1-unif-rotn}
 x_{kep} = \alpha^{-2/3} \, .
\end{equation}
The rotational speed at points in the pre-Keplerian region between
$X_{kep}$ and $X_{lim}$ will exceed the local Keplerian speed. Let
$X_{esc}$ be the point at which the rotational speed of the disk
reaches the escape velocity so that $x_{esc} = 2^{1/3} \,
\alpha^{-2/3}$. In some models, $X_{esc}$ will be located between
$X_{kep}$ and $X_{lim}$ and in other models the rotational speed
may not reach the escape speed between $X_{kep}$ and $X_{lim}$.
Let $X_{int}$ denote the nearer of $X_{esc}$ or $X_{lim}$ to the
point $X_{kep}$. Since the magnetic torque on the disk material is
negligible after the fill-up stage, we expect that the
super-Keplerian material in the region from $X_{kep}$ to $X_{int}$
will diffuse outwards into Keplerian orbits while conserving
angular momentum if viscosity is negligible. The material in the
region beyond $X_{int}$ will have sufficient kinetic energy to
flow away from the star.

When a quasi-steady Keplerian disk region is formed after
diffusion of the material in the pre-Keplerian region, let the
point $X_{end}$ be at its outer boundary. Then assuming that the
material at $X_{end}$ arrived from $X_{int}$ through angular
momentum conservation, we have $x_{end} = x_{int}^4 \, \alpha^2$.
Thus,
\begin{equation}\label{disk-kepl-xend-1}
x_{end} =   \cases{ 2^{4/3} \; \alpha^{-2/3}   & if $x_{lim} \geq
x_{esc}$\,,\cr x_{lim}^4 \; \alpha^2 & if $x_{lim} < x_{esc}$\,.
\cr }
\end{equation}

In order to form a disk region with Keplerian rotation we require
that $x_{lim} > x_{kep}$. When $x_{lim}$ is slightly larger than
$x_{kep}$ the Keplerian region will be in the form of a ring. For
larger values of $x_{lim}$ the radial extent of the Keplerian
region increases until $x_{lim} = x_{esc}$, which we refer to as
our \textit{optimal model}. When $x_{lim}$ increases beyond
$x_{esc}$, there is no increase in size of the Keplerian region.
The radial extent of the largest Keplerian disk region is given by
$\alpha^{-2/3} \leq x \leq 2^{4/3} \, \alpha^{-2/3}$.  If
viscosity plays a role during the process of diffusion of
super-Keplerian material into Keplerian orbits after fill-up, the
radial extent of the Keplerian region will be slightly different.
The presence of a weak viscous force may be helpful in the process
of readjustment of angular velocity.

We should point out that there are other situations in which
quasi-steady Keplerian disks with different radial extents can be
formed. For example, if the magnetic flux tube that assists in
forming the disk region is localized over the sector
$\Sigma_{\star, disk}$ such that the rotation speed at the
innermost point $X_{inn}$ of the pre-Keplerian disk region is
super-Keplerian, then the the inner radius of the quasi-steady
disk that is formed will be larger than $\alpha^{-2/3}R$.

\section{Constraints on the Magnetic field}\label{mag-fld-constraints}

\subsection{Magnetic Field at the Initial
Surface}\label{initial-Bm0-Bphi0} In this section we derive
expressions for $\mathcal{A}_{m, 0, lim}$ and $\mathcal{A}_{\phi,
0, lim}$ that are useful for finding solutions. In our models we
require that $\mathcal{A}_{m, 0} > 1$ at the points $P_0$ on the
initial surface. In the wind zone $W_{disk}$, the streamline from
$P_0$ travels to the point $Q$ on the shock surface and
$\mathcal{A}_m > 1$ at all points between $P_0$ and $Q$. In the
wind zone $W_{outer}$, the streamline from $P_0$ travels into
space and $\mathcal{A}_m > 1$ at all points between $P_0$ and
$P_{alf}$, which is the Alv\'en point on the streamline. In
Section \ref{azimuthal-wind-props}, we found that $B_\phi < 0$ and
${\mathcal A}_{\phi} < 0$ in the Northern Hemisphere in our
models. Equations (\ref{A-m-1}) and (\ref{A-phi-1}) show that
${\mathcal A}_{\phi}$ will be negative if $\mathcal{A}_m
> \ell/(r \sin\theta)$ because $\ell^2 > r^2 \, \sin^2\theta$ for
points between $P_0$ and $Q$, or between $P_0$ and $P_{alf}$.
Therefore, the field at the initial surface should satisfy the
condition $\mathcal{A}_{m, 0} > \ell/(r_0 \sin\theta_0)$, which
requires that $\mathcal{A}_{m, 0, lim}
>  y_{lim}$, where
\begin{equation}\label{A-m-0-lim-bound}
y_{lim} = \frac{x_{lim}}{x_0 \sin\theta_{0, lim}} \, .
\end{equation}
The dynamical equations governing the flow will also impose
constraints on the fields at the initial surface but we do not
consider them here. When the density distribution in the wind zone
is known, Equation (\ref{critical-rho}) gives
\begin{equation}\label{A-m-0-lim-value}
 \mathcal{A}_{m, 0, lim} = \left(\frac{\rho_{0, lim}}{\rho_{X, lim}}\right)^{{1/2}} \, .
\end{equation}
This is not an explicit equation for $\mathcal{A}_{m, 0, lim}$
because the value of $\rho_{X, lim}$ depends on the value of
$\mathcal{A}_{m, 0, lim}$. Equation (\ref{phi-eqn2}) gives a
steady state solution of the azimuthal equation of motion, which
leads to the expression for $\mathcal{A}_\phi$ in Equation
(\ref{A-phi-1}). When an appropriate value of $\mathcal{A}_{m, 0,
lim}$ has been determined, the value of $A_{\phi, 0, lim}$ is
given by
\begin{equation}\label{A-phi-0-lim-value}
 \mathcal{A}_{\phi, 0, lim} = \mathcal{A}_{m, 0, lim}
 \left(\frac{1 - y_{lim}^2}{\mathcal{A}_{m, 0, lim}^2 - y_{lim}^2}\right) \, .
\end{equation}
Equations (\ref{A-m-0-lim-value}) and (\ref{A-phi-0-lim-value})
are constraints that must be satisfied in a magnetic rotator wind
model in a steady state.

\subsection{Critical Values of the Surface Magnetic
Field}\label{interior-constraints} Our approach to Keplerian disk
formation does not depend on which theory of meridional
circulation speeds is valid and we consider circulation only as a
possible cause of some of the time-dependent behavior of disks.
The radial and azimuthal components of the surface magnetic fields
in rapidly rotating hot stars are subject to certain bounds
\citep{mah88, mah92}, which are
\begin{equation}\label{lowerbound}
B_{r, \star, L}=\left(4\pi \rho_{circ}\right)^{1/2}v_{circ} \left(
\frac{r_{circ}}{R}\right)^2
\end{equation}
 and
\begin{equation}\label{upperbound}
B_{\phi, \star, U}=\left[\frac{4\pi
GM(1-\Gamma)(1-\alpha^2)\rho_\star}{R}\right]^{1/2} \, .
\end{equation}
Subscript $circ$ denotes quantities relating to meridional
circulation. The effect of rotationally driven circulation on
meridional magnetic fields have been studied by \citet{mah69} and
by \citet{rob85}. The critical value given in Equation
(\ref{lowerbound}) is important for our models. If the radial
component $B_{r, \star}$ of the surface magnetic field is less
than $B_{r, \star, L}$, then meridional circulation currents
driven by rapid rotation can draw the meridional field lines
beneath the stellar surface in a time of order $t_{circ}=2\pi
r_{circ}/v_{circ}$. In this case, the meridional magnetic field
that provides the torque on the flow material will not persist
over periods of time that are much longer than $t_{circ}$.
However, if $t_{fill}$ is less than $t_{circ}$ a disk can form
before the field lines submerge. For circulation to be involved in
the time variation of disks, the meridional fields at the surface
should be strong enough to assist in the formation of disks but
weaker than $B_{r, \star, L}$. Also, a meridional component of the
field should be able to periodically emerge from the stellar
surface after being submerged by the circulation. The fields need
not cover the entire photosphere but emerge at least through the
sector $\Sigma_{\star, disk}$. The causes for the resurgence of
meridional field lines could be the buoyancy of magnetic flux
tubes or other MHD phenomena that occur in the stellar interior.
We know that the Sun displays such phenomena and it is reasonable
to expect that other stars do the same. The critical value given
by Equation (\ref{upperbound}) is used for ensuring that $B_{\phi,
\star}$ is less than $B_{\phi, \star, U}$ so that the sum of the
centrifugal force and magnetic force does not exceed gravity at
the photosphere. However, the magnetic fields in our models easily
satisfy this condition.

Meridional circulation velocities in rotating stars have been
discussed by several authors, e.g., \citet{swe50}, \citet{bak59},
\citet{mah68}, \citet{pav78}, \citet{tas95}. Unfortunately, no
circulation models are available for rapidly rotating hot stars
with winds. Sweet's first order expansion method gave relatively
modest values for circulation speeds near the stellar surface.
Maheswaran showed that the first order expansion method was
inadequate when the angular velocity is taken to be constant
because the term in $1/\rho$ does not appear in the first order
term for the circulation speed. In the case of uniform rotation,
the $1/\rho$ term appears in the second order and higher order
terms so that the circulation speeds are much larger than the
values estimated by Sweet. For nonuniform angular velocity
distributions, the $1/\rho$ factor in the circulation speed
appears in the first order term \citep{bak59, pav78}.
\citet{tas95} find that when eddy viscosity due to turbulence is
included, the $1/\rho$ factor disappears and circulation speeds
are similar to those obtained by \citet{swe50}. However, the
Tassoul \& Tassoul model is not appropriate for hot stars with
winds because it imposes the boundary condition that ${\bf v.n} =
0$, which is essentially the same as enforcing $v_{r, circ} = 0$
at the stellar surface. The fact that they obtain values of the
circulation speed that are smaller than the velocities for
mass-loss rates near the surface implies that their boundary
condition is not appropriate for hot stars with winds. Also, they
have not treated the surface region in detail so that it is not
clear whether the choking of circulation by eddy viscosity is a
consequence of the stringent boundary condition which requires the
flow to become transverse at the surface. The boundary condition
used by \citet{mah68} involves only the conservation of mass so
that $v_{r, circ}$ does not have to be zero near the photosphere.
In the absence of models with more appropriate boundary
conditions, we use the circulation velocity that is consistent
with the results of \citet{mah68} and \citet{pav78}, which is
\begin{equation} \label{vcirc1}
v_{circ} = \frac{LR^2}{GM^2} \, (1 - \Gamma)\alpha^2 \, \left\{1 +
{\frac{\bar{\rho}}{\rho_{circ}}} \, \left[(1 - \Gamma)\alpha^2 +
{\frac{|\Delta\Omega|}{\Omega}} \right] \right\} \, .
\end{equation}
Here, $\Omega$ is the angular velocity at the equator and
$\Delta\Omega$ is the change in angular velocity from equator to
pole. $\bar{\rho} = {3M}/({4\pi R^3})$ is the mean density of the
star. In the case of stars that are in almost uniform rotation we
can write the circulation velocity in the outer part of the
radiative diffusion zone in the form
\begin{equation} \label{vcirc2}
v_{circ} \approx \frac{3L(1 - \Gamma)^2 \alpha^4}{4 \pi
\rho_{circ} \, G M R}.
\end{equation}
We should point out that even if the Tassoul \& Tassoul approach
to meridional circulation is appropriate, the theory that we
propose for Keplerian disk formation is not affected. The only
difference is that meridional components of surface fields with
the strengths that we require will be able to survive for much
longer periods of time because the circulation speeds are very
small and the turnover time is long. In this case, rotationally
driven circulation will not be a factor in the time variation of
disks.

\subsection{ Time Scale for Ohmic Diffusion in the
Disk}\label{magnetic-field-diffusion} The azimuthal component
$\hat{B}_\phi$ of the magnetic field in the disk region will be in
opposite directions in opposite hemispheres and we wish to
determine whether there could be early decay of the field through
Ohmic diffusion because the height of the disk is small. The
diffusion time scale of the field is given by $t_{diff} = 4 \pi
\texttt{L}^2/\eta$, where $\texttt{L}$ is the length scale of the
field and $\eta$ is the magnetic diffusivity. Using the expression
given by \citet{spi62} for diffusivity transverse to the field, we
obtain $\eta = 1.42 \times 10^{14} T^{-3/2} \, {\textrm{cm}}^2 \,
\textrm{s}^{-1}$ for disk regions. Then, Equation
(\ref{fill-up-time-2}) for the disk fill-up time together with
$\texttt{L} = h_{fill}$, gives
\begin{equation}\label{diff-vs-fill-up}
\frac{t_{fill}}{t_{diff}} = \frac{1.13 \times 10^{13}\,
\hat{\rho}_{fill}}{r \,\rho v_n \, \hat{T}^{3/2} \,
\tan\delta_{fill}}\, .
\end{equation}
We find that the diffusion times for the fields in the disk
regions of our models are very much longer than the fill-up times
so that decay due to Ohmic diffusion will not be a factor.

\section{Dipole-type Magnetic Fields in Wind Zones}\label{mag-fld-models}
Since we wish to study the formation and properties of Keplerian
disks, we consider models in which meridional fields have a
dipole-type structure. In particular, we look for solutions with
stream functions that have the form given by
\begin{equation}\label{streamlines-1}
\psi(r,\theta) \equiv \frac{\sin^2 \theta}{r^\lambda} =
\frac{\sin^2 \theta_0}{r_0^\lambda} \, ,
\end{equation}
where $\lambda$ is a constant. The field lines that coincide with
streamlines given by this equation correspond to a dipole-type
field with
\begin{equation}\label{br-bth-1}
B_r = \frac{2C}{r^{2+\lambda}}\cos\theta \quad {\rm and} \quad
B_\theta = \frac{\lambda C}{r^{2+\lambda}}\sin\theta \, ,
\end{equation}
where $C$ is a constant. When $\lambda = 1$ this gives a regular
dipole field. However, even if the magnetic field at the stellar
surface is that of a regular dipole, the field lines in the wind
zones may be slightly modified by the flow. In realistic models it
is likely that the values of $\lambda$ will be smaller than 1.
When $\lambda$ approaches $0$ the transverse component $B_\theta$
approaches $0$ and when $\lambda = 0$ the field becomes purely
radial. We consider models in which $0 < \lambda \leq 1$.

In our applications, the disk has a shock boundary given by
$\theta = (\pi/2) - \delta$ in the Northern Hemisphere. If the
meridional streamline from the point $P_0(r_0, \theta_0, 0)$ on
the initial surface arrives at the point $Q(r_Q, (\pi/2) -\delta,
0)$ on the shock boundary, we have
\begin{equation}\label{r-q-eqn}
r_Q = r_0 \left(\frac{\cos \delta}{\sin
\theta_0}\right)^{2/\lambda} \; .
\end{equation}
As described in Section \ref{rotation-of-disk-region}, we take
$X(r_X, \pi/2, 0)$ to be the point on the equatorial plane that is
associated with the point Q such that $r_X = r_Q \cos\delta$.
Equation (\ref{r-q-eqn}) gives the relationship between $\theta_0$
and $x$ in the form
\begin{equation}\label{x-theta-0-eqn}
\sin^2 \theta_0 = \left(\frac{x_0}{x}\right)^\lambda\cos^{2 +
\lambda} \delta \; .
\end{equation}
The initial surface in our models is close enough to the stellar
surface so that we can take $r_0 \approx R$. Also, the disk is
sufficiently thin such that $\cos\delta \approx 1$. Then, Equation
(\ref{x-theta-0-eqn}) can be written as
\begin{equation}\label{streamlines-2}
\sin \theta_0 = \frac{1}{x^{\lambda/2}} \, .
\end{equation}
We can construct models by prescribing the values of any two of
the quantities $x_{lim}$, $\theta_{0,lim}$ and $\lambda$. Then,
Equation (\ref{streamlines-2}) fixes the value of the third
quantity. In models where $\theta_{0,lim}$ and $x_{lim}$ are
known, we obtain
\begin{equation}\label{lambda}
\lambda = - \frac{2 \ln (\sin\theta_{0, lim})}{\ln (x_{lim})}.
\end{equation}
Equations (\ref{br-bth-1}) and (\ref{x-theta-0-eqn}) yield
\begin{equation}\label{bm-Q-x}
B_{m,Q} = B_{m, 0} \left\{ \frac{\lambda^2}
{x^{4+\lambda}\left[4\left(x^\lambda - 1\right) +\lambda^2
\right]} \right\}^{{1/2}} \, .
\end{equation}
We assume that over the sector $\Sigma_{0, disk}$ of the initial
surface, the density, $\rho_0$, and meridional speed of the flow,
$v_{m, 0}$, are constant. Then, using $B_{m, 0} =
\sqrt{4\pi\rho_0}\; v_{m, 0}\,\mathcal{A}_{m, 0}$ we obtain
\begin{equation}\label{Am-0-Am-0-lim}
\mathcal{A}_{m,0} = \mathcal{A}_{m, 0, lim} \left\{
\frac{x_{lim}^\lambda\left[4\left(x^\lambda -1\right) + \lambda^2
\right]}{x^\lambda\left[4\left(x_{lim}^\lambda -1\right)
+\lambda^2 \right]}\right\}^{{1/2}} \, .
\end{equation}
For our dipole-type fields, we obtain $y_{lim} =
x_{lim}^{({2+\lambda}/{2})}$ for the quantity defined in Equation
(\ref{A-m-0-lim-bound}). This expression can be substituted in
Equation (\ref{A-phi-0-lim-value}) to compute the value of
$\mathcal{A}_{\phi, 0, lim}$ in terms of $x_{lim}$ and
$\mathcal{A}_{m, 0, lim}$.

\section{Models and Applications}\label{models}
If we know the density, velocity and magnetic field strength at
the stellar surface, we can use our system of equations and
boundary conditions to compute solutions for the wind zones and
the disk region. Because the flow streamlines and the magnetic
field lines coincide in a meridional plane, a convenient approach
is to introduce a meridional stream function $\psi (r, \theta)$
for the wind zone and obtain solutions for it in terms of known
values at the initial surface. In this paper we do not attempt to
completely solve our system of equations. For the present, we
focus on the formation and properties of Keplerian disks. We
derive two results that help us develop a method to pursue this
goal. For given values of $\theta_0$ and $\mathcal{A}_{\phi, 0}$,
Equation (\ref{critical-l2}) gives $d\ell/d\mathcal{A}_{m, 0}
> 0$, which means that when $B_{m, 0}$ increases so does
$x_{lim}$. Also, for given values of $\theta_0$ and
$\mathcal{A}_{m, 0}$, we obtain $d\ell/d\mathcal{A}_{\phi, 0} <
0$. Since $\mathcal{A}_{\phi, 0} < 0$, when $|B_{\phi, 0}|$
increases $x_{lim}$ also increases.

We consider models in which $\psi (r, \theta)$ has the form given
by Equation (\ref{streamlines-1}) and compute solutions for a
large number of different cases. First, we select a value for the
uniform rotation rate, $\alpha$, of the sector $\Sigma_{\star,
disk}$. Then, we select a value for $\theta_{0, lim}$ from a range
of plausible values. This specifies the location of the point
$P_{0, lim}$, which is the upper boundary of the sector
$\Sigma_{0, disk}$ on the initial surface. Next, we choose an
appropriate value of $x_{lim}$, which fixes the magnitudes of
$B_{m, 0, lim}$ and $B_{\phi, 0, lim}$ at $P_{0, lim}$. In our
models, the minimum value of $x_{lim}$ that is required for the
formation of a quasi-steady Keplerian disk is given by $x_{lim} =
x_{kep} = \alpha^{-2/3}$. Let $B_{m, 0, min}$ denote the value of
$B_{m, 0, lim}$ in this case. Then, $B_{m, 0, min}$ is the minimum
meridional magnetic field that is required at the initial surface
for Keplerian disk formation. When we increase the value of
$x_{lim}$, the pre-Keplerian region between $X_{kep}$ and
$X_{lim}$ increases in size as does the radial extent of the
quasi-steady Keplerian disk region into which it evolves. This
continues until the value of $x_{lim}$ reaches the value
$x_{esc}$. Let $B_{m, 0, opt}$ be the value of of $B_{m, 0, lim}$
when $x_{lim} = x_{esc} =  2^{1/3} \, \alpha^{-2/3}$. For values
of $x_{lim}$ larger than $x_{esc}$, although the pre-Keplerian
region is larger, the radial extent of the corresponding
quasi-steady Keplerian disk region does not increase in size
because only the portion of the pre-Keplerian region between
$x_{lim}$ and $x_{esc}$ evolves into the quasi-steady Keplerian
disk region. Thus, $B_{m, 0, opt}$ is smallest meridional field
for which the quasi-steady Keplerian disk has the largest radial
extent. The model with $B_{m, 0, lim} = B_{m, 0, opt}$ is our
optimal model. We denote the corresponding azimuthal field by
$B_{\phi, 0, opt}$. For values of $B_{m,0, lim}
> B_{m, 0, opt}$ the radial extent of the quasi-steady Keplerian
region is the same as that of the optimal model. Let $P_{0,
kep}(r_0, \theta_{0, kep}, 0)$ be the point on the initial surface
that corresponds to $X_{kep}$. Then, the sector $\Sigma_{0, disk}$
is given by $\theta_{0, kep} < \theta_0 < \theta_{0, lim}$.

In applications to models, we take the initial surface,
$\Sigma_0$, to be the sonic surface. Then, $v_{m, 0} = c_s$, where
$c_S^2 = \Re \, T/\mu$. Here, $\Re$ is the gas constant and $\mu$
is the mean molecular weight. Since this surface is close enough
to the stellar surface we take $r_0 \approx R$. We consider a thin
disk region with conical shock boundaries given by $\theta =
(\pi/2) - \delta$ and $\theta = (\pi/2) + \delta$ during the
fill-up stage, where $\delta$ is the same for all points on the
shock surface. At the end of the fill-up stage $\delta =
\delta_{fill}$. Fill-up occurs on a time of order $t_{fill}$,
which is much longer than the flow time $t_{flow} = R/v_\theta$
near the shock surface. The magnetic field emerging from the
sector $\Sigma_{0, disk}$ has a dipole-type structure given by
Equations (\ref{br-bth-1}). Since the values of $\theta_{0, lim}$
and $x_{lim}$ have been chosen, Equation (\ref{lambda}) fixes the
value of $\lambda$. We assume that the initial density, $\rho_0$,
is constant over $\Sigma_0$ and we use the mass loss rate,
$\dot{M}$, to obtain
\begin{equation}\label{rho-0-comp}
\rho_0 = \frac{\dot{M}}{4\pi \, r_0^2 c_s} \, .
\end{equation}
Using the definition of Alfv\'en velocity and the values of
azimuthal quantities given in Sections \ref{alfven-velocities} and
\ref{azimuthal-wind-props}, we can write
\begin{equation}\label{B-m0-value}
 B_{m, 0} =  \mathcal{A}_{m,
 0}\left(\frac{\Re\,T}{\mu}\right)^{1/4}\left(\frac{\dot{M}^{1/2}}{r_0}\right)
\end{equation}
and
\begin{equation}\label{B-phi0-value}
 B_{\phi, 0} =  \left(\frac{\mathcal{A}_{\phi,0} \, \mathcal{A}_{m,0}}
{\mathcal{A}_{m,0} - \mathcal{A}_{\phi, 0}}\right)
\left(\frac{\mu}{\Re\,T}\right)^{1/4} \left(\frac{\dot{M}^{1/2}}{
R}\right) v_{\phi, \star} \, ,
\end{equation}
where $v_{\phi, \star}$ is the rotational speed at the point
$P_\star$ of the photosphere and is given by Equations
(\ref{v-phi-star-eqn}) and (\ref{omega-max}) with $\alpha_\star =
\alpha$ and $\theta_\star = \theta_0$. In general,
$\mathcal{A}_{m,0}$ and $\mathcal{A}_{\phi, 0}$ are functions of
colatitude $\theta_0$.

In order to proceed further, we need to determine either the
meridional velocity, $v_m$, or the density, $\rho$, of the flow as
it approaches the shock boundary of the disk. When one of these
quantities is known the other can be computed using Equation
(\ref{vm-bm-1}) in models with dipole-type fields. In a study
where detailed properties of the wind zones are required, we
should convert the Bernoulli Equation (\ref{bernoulli}) to an
equation in $v_m$ as done by \citet{mes68} or to an equation in
$\rho$ as in \citet{sak85}. Both forms of the equation are quite
complicated and the numerical methods that must be used to find
solutions are too laborious for our immediate needs. Since we
require values of $v_m$ for the flow only near the shock surface,
we use the $\beta$-law for the velocity of stellar winds and
assume that
\begin{equation}\label{v-m-Q}
v_{m, Q} = v_\infty \, \left(1 - \frac{R}{r_Q}\right)^\beta \, ,
\end{equation}
where $\beta$ is a constant such that $0.5 < \beta < 2$ (e.g.,
Lamers \& Cassinelli 1999). We use Equation (\ref{v-m-Q}) to
compute the flow velocity adjacent to the point $X_{lim}$. Then,
we compute the density at that point by using Equations
(\ref{vm-bm-1}) and (\ref{bm-Q-x}). Next, we use the equations in
Section \ref{initial-Bm0-Bphi0} to obtain the values of
$\mathcal{A}_{m, 0, lim}$ and $\mathcal{A}_{\phi, 0, lim}$, and
then use Equation (\ref{Am-0-Am-0-lim}) to find the value of
$\mathcal{A}_{m, 0}$. At this stage we have the value of
$\mathcal{A}_{\phi, 0, lim}$ but because we have not specified the
azimuthal field along the initial surface, we do not have any
conditions on how $\mathcal{A}_{\phi, 0}$ relates to
$\mathcal{A}_{\phi, 0, lim}$, or how $B_{\phi, 0}$ relates to
$B_{\phi, 0, lim}$. In our models, we choose the azimuthal field
at the initial surface such that $B_{\phi, 0}$ is directly
proportional to $v_{\phi, 0}$ at points on the sector $\Sigma_{0,
disk}$. Then, $\mathcal{A}_{\phi, 0}$ is constant over this sector
so that its value is equal to $\mathcal{A}_{\phi, 0, lim}$. When
$\mathcal{A}_{m, 0}$ and $\mathcal{A}_{\phi, 0}$ are known,
Equations (\ref{B-m0-value}) and (\ref{B-phi0-value}) give the
values of $B_{m, 0}$ and $B_{\phi, 0}$.

\placetable{table-star-properties}

\section{Results}\label{results}
We apply the method of solution described in Section \ref{models}
to stellar models whose basic properties are given in Table
\ref{table-star-properties}. The values of the density, and the
flow velocity at the initial surface are also shown in this table.
We consider a variety of different cases by taking different
combinations of values of $\alpha$, $\theta_{0, lim}$, $x_{lim}$,
and $\beta$. Table \ref{table-stars-alpha-max} gives the upper
bounds on rotation rate, $\alpha$, for which quasi-steady
Keplerian disks can be formed when the magnetic field components
have their optimal values. We see that the upper bound on $\alpha$
for Keplerian disk formation decreases when the terminal velocity,
$v_\infty$, of the wind decreases and this can be explained in
terms of the necessary condition (\ref{disk-formation-condn}) that
we derive in Section \ref{condn-for-pre-kepl-disk}, where the
magnetic pressure in the disk plays a role even though the
magnetic force is small compared with gravity or centrifugal
force. The upper bounds on the rotation rate, $\alpha$, for
Keplerian disk formation will be larger for stars in which the
normal component of the flow velocity into the shock surface is
larger than what we have assumed in our models.

\placetable{table-stars-alpha-max}
\placetable{table-stars-Bcritical}

Table \ref{table-stars-Bcritical} displays the values of the
critical quantities $B_{m, 0, min}$, $B_{m, 0, opt}$ and $B_{\phi,
0, opt}$ for the magnetic field at the initial surface for models
with selected values of $\alpha$ when $\theta_{0, lim} = 65^\circ$
and $\theta_{0, lim} = 75^\circ$. We use Equation
(\ref{fill-up-time-2}) to compute the fill-up times shown in the
table. The table also shows the values of $B_{r, \star, L}$, which
is the minimum field strength required by the meridional component
of the field at the photosphere to withstand the effects of
rotationally driven circulation. We find that the values of $B_{m,
0, min}$ and $B_{m, 0, opt}$ are about 30 G and 70 G,
respectively, for a B0 star and about 1 G and 1.6 G for a B9 star.
The magnitudes of $B_{\phi, 0, opt}$ are about 15 G and 0.2 G for
the same stars. Except in models of the B9 star, we find that the
values of $B_{m, 0, opt}$ are less than $B_{r, \star, L}$. In the
B9 models, they are of the same order. In a rapidly rotating O3
star, $B_{m, 0, min}$ is about 500 G and $B_{m, 0, opt}$ is about
1000 G. In an 06.5 star the corresponding values are about 100 G
and 200 G, respectively. In models of O3 stars, the values of
$B_{m, 0}$ required to form Keplerian disks are larger than $B_{r,
\star, L}$ and in O6.5 stars they are of the same order. At the
end of the fill-up stage, the density of the disks vary from about
$5 \times 10^{-13} \textrm{gm cm}^{-3}$ for B0 stars to about $7
\times 10^{-17} \textrm{gm cm}^{-3}$ for B9 stars. The values for
O3 and O6.5 stars are, respectively, of order $10^{-11} \textrm{gm
cm}^{-3}$ and $10^{-12} \textrm{gm cm}^{-3}$. The values of the
disk density $\hat{\rho}$ depend on the values of $\dot{M}$ and
$v_{\infty}$. If the mass loss rates near the equator for the
different stellar models are higher than the values shown in Table
\ref{table-star-properties}, the disk densities will be larger
than the values shown in Table \ref{table-stars-Bcritical}. The
disk fill-up times have been computed for models with
$\delta_{fill} = 10^\circ$ and we find that these times are of the
order of a few months for models of O-type stars and about a week
for B9 models.

\placetable{table-stars-Fmag-Fvisc}

Table \ref{table-stars-Fmag-Fvisc} shows the average values of the
ratio of the magnetic force to centrifugal force and the ratio of
the viscous force to magnetic force in the disk regions after the
fill-up stage. Because of the increased density, the magnetic
force components ${\mathcal{F}}_r$ and ${\mathcal{F}}_{\phi}$ in
the disk region, given by Equations (\ref{disk-mag-force-r}) and
(\ref{disk-mag-force-phi}), become small compared with the
centrifugal force or gravity. The results given in  Table
\ref{table-B2-Bcrit-alpha} for the optimal models of the B2 star
show that when $\Sigma_{\star, disk}$ is located near the latitude
of $10^\circ$, its spread in latitude is only about $2^\circ$ and
when it is located near the latitude of $25^\circ$ the spread is
about $5^\circ$. The spread in latitude required for the sector
$\Sigma_{\star, disk}$ in optimal models is smaller when it is
closer to the equator. Table \ref{table-B2-Bm0-Bphi0-xlim}
displays values of $B_{m, 0, lim}$ and $B_{\phi, 0, lim}$ for
different values of $x_{lim}$ in models of the B2 star with
$\alpha = 0.5$ and $\theta_{0, lim}=70^\circ$. This verifies the
result we derived in in Section \ref{models} that the magnitudes
of $B_{m, 0}$ and $B_{\phi, 0}$ increase when the values of
$x_{lim}$ are increased.

\placetable{table-B2-Bcrit-alpha}
\placetable{table-B2-Bm0-Bphi0-xlim}

\section{Discussion}\label{discussion}
In Section \ref{rotn-theorem}, we establish an important theorem
connecting the angular velocity at points on the photosphere with
the angular velocities at corresponding points in a disk region
with no meridional motion, when the two sets of points are linked
by field lines of a magnetic field of sufficient strength and the
system is in a steady state with axial symmetry. This theorem is
applicable in astrophysical situations in which a central rotating
object is linked to an equatorial disk region by a magnetic field
with the required strength and structure. In this paper, we use
the theorem to discuss the formation of Keplerian disk regions
around magnetic rotator stars. If the sector $\Sigma_{\star,
disk}$ of the photosphere rotates with approximately uniform
angular velocity and is linked to a pre-Keplerian disk region by a
dipole-type magnetic field of sufficient strength, then during the
fill-up stage the disk region will have the same uniform angular
velocity. In Section \ref{condn-for-pre-kepl-disk} we derive a
necessary condition for the formation of a Keplerian disk. It
requires that the normal component of the flow velocity near the
shock boundary must be sufficiently large for Keplerian disks to
form when a magnetic field is present and when the rotation rate
is high. This constraint results from the fact that magnetic
pressure in the disk region plays a role in inhibiting Keplerian
disk formation even when the magnetic force is small compared with
gravity or centrifugal force.

Our models require modest magnetic fields such that the meridional
Alfv\'en speeds are larger than the speeds of the outflow in the
wind zone $W_{disk}$ between the stellar surface and the shock
surface. Several authors \citep{bab97b,don01,mac03} have referred
to observational evidence for the presence of surface magnetic
fields in hot stars with strengths of several hundred gauss. We
know that fossil fields can be present in radiative envelopes of
early type stars. \citet{spi62} gives the Ohmic decay time for a
stellar field with length scale $\texttt{L} = R$ and appropriate
resistivity to be $2 \times 10^{-13} \, T^{3/2} \, R^2$ s. If we
take $T = T_{eff}$, we see that this time is much longer than the
evolutionary time for all the stars except for the B9 star. For
the B9 star they are of the same order. In the radiative envelopes
of the stars, $T$ is much larger than $T_{eff}$ and the decay time
will be much longer for all the stars. Recently, \citet{cha01}
showed that magnetic fields may be generated through dynamo action
at the core-envelope boundary of a hot star and \citet{mac03} have
found that such a field can be transported to the surface region
in a time shorter than an evolutionary time of the star. In view
of the observational evidence and theoretical support it is
reasonable for us to consider models with modest surface fields.

Our approach to modeling the formation of Keplerian disks consists
of a disk fill-up process and a process of adjustment of
super-Keplerian disk material into Keplerian orbits. Although the
second process may begin during the later stages of the first
process, it is reasonable to consider them separately because
initially the density in the disk region is small and the magnetic
field plays a role in the rotation of the disk material. Towards
the end of the fill-up stage, the density of the disk region
increases to much larger values and the magnetic force in the disk
becomes negligible when compared with centrifugal force or
gravity. This allows the super-Keplerian material in the
pre-Keplerian disk region to move into Keplerian orbits while
conserving angular momentum when there is no viscous force.
Although viscosity is negligible at the end of the fill-up stage,
the onset of a weak viscous force during the process of diffusion
of super-Keplerian material into into Keplerian orbits will be
helpful.

When the meridional field $B_{m, 0}$ at the initial surface is
only slightly larger than the minimum value $B_{m, 0, min}$ that
is required for disk formation, the Keplerian region will be in
the shape of a ring. For increasing values of $B_{m, 0}$ the
radial extent of the Keplerian region increases and the region
reaches its largest extent when $B_{m, 0}$ equals the optimal
value $B_{m, 0, opt}$. For values of $B_{m, 0}$ larger than $B_{m,
0, opt}$ there is no increase in the radial extent of the
quasi-steady Keplerian region. When the sector $\Sigma_{\star,
disk}$ of the photosphere has uniform rotation, the inner and
outer radii of the maximal quasi-steady Keplerian region of the
disk are given by $\alpha^{-2/3}R$ and $2^{4/3}\alpha^{-2/3}R$,
respectively, where $R$ is the stellar radius and $\alpha$ is the
ratio of the rotational speed to the Keplerian speed at
$\Sigma_{\star, disk}$. While the magnetic field strength at the
stellar surface determines the extent of a quasi-steady Keplerian
disk region, we see that both the inner and outer radii of a
maximal disk are given in terms of $\alpha$. The inner radius is
the distance at which there is corotation with Keplerian rotation
speed. The outer radius of the maximal disk is determined by the
point where the escape velocity is reached in a pre-Keplerian
region and it is larger for smaller values of $\alpha$. Our result
for the maximal radial extent of a quasi-steady Keplerian region
of the disk is obtained without including the effect of viscous
forces. As shown in Table \ref{table-stars-Fmag-Fvisc}, the
viscous force in the disk region is negligible at the end of the
fill-up stage. If a weak viscous force plays a role during the
transition from a pre-Keplerian to a quasi-steady Kelperian disk
region, the radial extent of the quasi-steady region will be
slightly different. There are other situations that we do not
discuss here in which quasi-steady Keplerian disks of different
radial extent can be formed. For example, if the magnetic field
lines from the star to the disk region are confined to a sector
$\Sigma_{\star, disk}$ such that the rotation speed at the
innermost point of the pre-Keplerian disk region is
super-Keplerian, then the inner radius will be larger than
$\alpha^{-2/3}R$. Thus, if the rotation rate and the extent of a
Keplerian disk region of a star can be obtained from observations,
our models can be used to make some deductions about the magnetic
field of the star.

The minimum and optimal meridional magnetic fields required for
the formation of Keplerian disks in models of B-type stars are,
respectively, of order 1 G to 10 G. These fields are weaker than
$B_{r, \star, L}$, which is the lower bound derived by
\citet{mah88, mah92} for the meridional field at the stellar
surface to withstand the effects of meridional circulation. The
fill-up times in models of all the different stars that we
consider in this paper are shorter than the corresponding
circulation times. Thus, although meridional circulation may not
affect the formation of disks in B-type stars, its effect on the
stellar field can influence some observed properties of disks.  On
the other hand, the situation in regard to O-type stars is
different. In models of O6.5 stars, meridional magnetic fields
required for Keplerian disk formation are of the order of 100 G,
which is of the same order as $B_{r, \star, L}$, and in O3 stars
they are about 1000 G, which is much larger than $B_{r, \star,
L}$. Although the fill-up times in O-type stars are shorter than
the meridional circulation times, if these stars posses the
meridional fields with the strengths that are required for
Keplerian disk formation, circulation may not be involved in
causing any observable changes in disks. Our numerical results
indicate that for reasonable values of the rotation rate and
surface magnetic field strengths, B-type stars are likely to
possess Keplerian disks. This is consistent with the results for
the distribution of Be stars obtained by \citet{pol91}.

Our models do not require that the meridional field, $B_{m, 0}$,
be uniformly strong across the stellar surface for the formation
of a Keplerian disk. The field must have the required strength on
the sector $\Sigma_{0, disk}$ and the rotation should be
approximately uniform on the sector $\Sigma_{\star, disk}$. The
entire star need not be in uniform rotation. For all reasonable
locations of $\Sigma_{\star, disk}$ in all our models, we find
that its spread in latitude needs to be only a few degrees. We
should emphasize that the assumption of axial symmetry has been
made only to simplify the mathematics. Our model for disk
formation does not depend on an aligned dipole field or even an
axially symmetric field. Our results will qualitatively apply to
oblique rotator models or to stars with magnetic fields consisting
of flux loops that emerge from lower latitudes and thread the disk
region.

The strengths of the meridional fields required for Keplerian disk
formation in our models suggest that a possible scenario for the
variation with time of properties of Keplerian disks in some
B-type stars is the following. These stars possess magnetic fields
that are mainly sub-photospheric and from time to time, due to
magnetic buoyancy or other interior phenomena, meridional magnetic
field lines emerge from the stellar surface. The strength of the
meridional component of the field satisfies the condition $B_{m,
0, min} < B_{m, 0} < B_{r, \star, L}$ so that it is able to assist
in the formation of Keplerian disks. After the formation of a
disk, the meridional field lines are drawn below the stellar
surface in a time scale of the order of the circulation time,
which is longer than the disk fill-up time, $t_{fill}$. This shuts
off the supply of magnetically torqued material to the disk from
non-equatorial latitudes. The resurgence and submersion of
meridional fields across the stellar surface could explain some of
the variations of the disk region over different time scales. We
should emphasize that the theory that we propose for Keplerian
disk formation does not depend on which theory of meridional
circulation speeds is valid. We consider circulation only as a
possible cause of variation of disks with time. Thus, the Be star
phenomenon may be an interesting example where the effects of
rotation, circulation and magnetic activity play a significant
role. Obviously, stars in which $B_{m, 0, min}
> B_{r, \star, L}$ will not exhibit this aspect of time variation.
It is beyond the scope of the present work to consider details of
the time dependent behavior of the surface field caused by
rotation, circulation and outflow or the depletion of flux due to
instabilities. Obviously, such a study will be greatly beneficial
for a better understanding the role of magnetic fields in the
formation, properties and variability of disks.

Some of the basic goals of this paper are to understand how
magnetic fields of rotating stars can influence the formation of
Keplerian disks and to explain the time variability observed in Be
stars. The MTD model of \citet{cas02} has the same objectives and
the flow regions that we consider in our models are similar to
those in the MTD model. One of the main differences between our
models and the MTD model is that we are able to establish a result
for the angular velocity distribution during the fill-up stage of
a disk. The MTD model uses an empirical formula for the rotational
velocity in the disk region, which may be appropriate in WCD
models. A second difference is that the meridional fields in our
models can be weaker than the critical strength, $B_{r, \star, L}$
and yet be strong enough to provide the torque required by the
flow material to reach rotational speeds that lead to the
formation of Keplerian disks. The MTD model requires stronger
fields because the torque is applied to the disk and, therefore,
their strengths are larger than $B_{r, \star, L}$. As we show in
this paper, a problem with stronger fields in the disk region is
that they can inhibit the formation of a Keplerian region. An
important aspect of our models is that for the strengths of fields
that we consider, the magnetic force in the disk region becomes
negligible at the end of the fill-up stage. Thus, the one-armed
spiral pattern of the Global Disk Oscillation model \citep{oka97}
can be used to explain the V/R variability in disks of Be stars.

\acknowledgements I wish to thank J. P. Cassinelli for valuable
discussions and comments on the manuscript and gratefully
acknowledge a University of Wisconsin Marathon County Foundation
Summer Grant.

\appendix

\section{Equations for Azimuthal Components and Alfv\'en Velocities}\label{apndx-1}
The value of the quantity $\ell$ for any streamline in the
Northern Hemisphere is given by
\begin{equation}\label{critical-l2}
\ell = r_0 \sin \theta_0 \left[\frac{{\mathcal
A}_{m,0}\left(1-{\mathcal A}_{m,0} \, {\mathcal
A}_{\phi,0}\right)} {{\mathcal A}_{m,0}-{\mathcal A}_{\phi,0}}
\right]^{1/2} = r \sin \theta \left[\frac{{\mathcal A}_m
\left(1-{\mathcal A}_m \, {\mathcal A}_\phi\right)} {{\mathcal
A}_m-{\mathcal A}_\phi} \right]^{1/2}\, .
\end{equation}
We derive this by first substituting for $\mathcal{L}$ from
Equation (\ref{critical-l1}) in Equation (\ref{phi-eqn2}) and then
eliminating $\omega$ between this new equation and Equation
(\ref{v-phi-1}). Next, we substitute for $\xi$ and $\chi$ from
Equation (\ref{xi-eqn})and use Equation (\ref{alfven-mach-defn})
to replace $B_m$ and $B_\phi$ by $\mathcal{A}_m$ and
$\mathcal{A}_\phi$.

We can write an expression for $v_\phi$ and $B_\phi$ in the
Northern Hemisphere in the forms
\begin{equation}\label{v-phi-3}
v_\phi = \omega r \sin\theta
\left(\frac{\mathcal{A}_m}{\mathcal{A}_m - \mathcal{A}_\phi}
\right)
\end{equation}
and
\begin{equation}\label{b-phi-2}
B_\phi = \omega r \sin\theta \left(\frac{4\pi \rho_0
v_{m,0}}{B_{m,0}}\right) \left(\frac{\mathcal{A}_m^2 \,
\mathcal{A}_\phi}{\mathcal{A}_m - \mathcal{A}_\phi} \right)
 \, .
\end{equation}
Along an open streamline, both $v_\phi$ and $B_\phi$ have
removable singularities at $r \sin\theta = \ell$ where $\rho =
\rho_c$.  The values of $\rho_c$ and $\ell$ to be inserted in
these equations are given by Equations (\ref{critical-rho}) and
(\ref{critical-l2}), respectively. Expressions for ${\mathcal
A}_{m}$ and ${\mathcal A}_{\phi}$ in the Northern Hemisphere are
obtained from Equations (\ref{vm-bm-1}), (\ref{critical-rho}),
(\ref{v-phi-2}) and (\ref{b-phi-1}). These are
\begin{equation}\label{A-m-1}
{\mathcal A}_{m} = \left(\frac{\rho}{\rho_c}\right)^{1/2}
\end{equation}
and
\begin{equation}\label{A-phi-1}
{\mathcal A}_{\phi} = \left(\frac{\rho_c}{\rho}\right)^{1/2} \,
\left[ \frac{1 - \{\ell^2/( r^2 \sin^2\theta )\}}{1 - (\rho_c
/\rho)\{\ell^2/( r^2 \sin^2\theta )\}} \right]
\end{equation}
at points in all wind zones in the Northern Hemisphere.

{}

\clearpage

\clearpage

\begin{deluxetable}{lrrrccrrcc}
\tabletypesize{\scriptsize} \tablecaption{Properties of the
different stellar models considered \label{table-star-properties}}
\tablewidth{0pt} \tablecolumns{2} \tablehead{\colhead{Spectral}
&\colhead{$M$} &\colhead{$R$} &\colhead{$L$} &\colhead{$T_{eff}$}
 &\colhead{${\dot M}$} &\colhead{$v_\infty$} &\colhead{$\rho_0$}
&\colhead{$v_{m,0}$} &\colhead{${v_{\phi,\star,eq,crit}}$} \\
\colhead{Type} &\colhead{($M_\odot$)} &\colhead{($R_\odot$)}
&\colhead{($L_\odot$)} &\colhead{($10^4 \,^\circ$K)}
&\colhead{($10^{-9} M_\odot \textrm{yr}^{-1}$)}
&\colhead{(${{\textrm {km s}}^{-1}}$)} &\colhead{($\textrm{gm
cm}^{-3}$)} &\colhead{($10^6\, \textrm{cm s}^{-1}$)}
&\colhead{($\textrm{km s}^{-1}$)}} \startdata
O3   & $55\;\;$  & $14\;\;$ & $1.10\times 10^6$ & $5.0$  & $9100$    & $3100$ & $1.85 \times 10^{-11}$ & $2.60$ & $592$ \\
O6.5 & $29\;\;$  & $10\;\;$ & $2.31\times 10^5$ & $4.0$  &  $310$    & $2500$ & $1.38 \times 10^{-12}$ & $2.32$ & $661$  \\
B0   & $15\;\;$  & $6.6$    & $4.12\times 10^4$ & $3.2$  &   $27$    & $1300$ & $3.09 \times 10^{-13}$ & $2.08$ & $634$  \\
B2   & $8.3$     & $4.5$    & $  5110$          & $2.3$  &  $0.4$    & $ 840$ & $1.16 \times 10^{-14}$ & $1.76$ & $589$  \\
B5   & $4.5$     & $3.5$    & $   560$          & $1.5$  &  $0.01$   & $ 580$ & $5.95 \times 10^{-16}$ & $1.42$ & $495$ \\
B9   & $2.6$     & $2.6$    & $    61$          & $1.0$  &  $0.0013$ & $ 460$ & $1.72 \times 10^{-16}$ & $1.16$ & $437$ \\
\enddata
\tablecomments{Entries for the basic stellar properties are from
\citet{bjo93}. We use $L/L_\odot = (R/R_\odot)^2
\,(T_{eff}/T_{\odot, eff})^4 $ to compute luminosity.}
\end{deluxetable}

\clearpage

\begin{deluxetable}{lccc}
\tabletypesize{\scriptsize} \tablecaption{Upper bounds on the
rotation rate $\alpha$ \label{table-stars-alpha-max}}
\tablewidth{0pt} \tablecolumns{4} \tablehead{ \colhead{Spectral}
&\colhead{$\beta = 0.75$}
&\colhead{$\beta = 1.00$} &\colhead{$\beta = 1.50$} \\
\colhead{Type} &\colhead{$\alpha$} &\colhead{$\alpha$}
&\colhead{$\alpha$}} \startdata
O3   & $0.91$ &$0.86$ & $0.72$   \\
O6.5 & $0.89$ &$0.80$ & $0.65$   \\
B0   & $0.76$ &$0.65$ & $0.50$   \\
B2   & $0.65$ &$0.55$ & $0.41$   \\
B5   & $0.59$ &$0.49$ & $0.36$   \\
B9   & $0.55$ &$0.45$ & $0.33$   \\
\enddata
\tablecomments{Values shown are for optimal models of Keplerian
disks when $\theta_{0, lim} = 70^{\circ}$.}
\end{deluxetable}

\clearpage

\begin{deluxetable}{lccccccccrlc}
\tabletypesize{\scriptsize} \tablecaption{Critical magnetic field
strengths, disk fill-up times and time scales for circulation
\label{table-stars-Bcritical}} \tablewidth{0pt} \tablecolumns{2}
\tablehead{\colhead{Spectral} &\colhead{$\alpha$}
&\colhead{$\theta_{0, lim}$} &\colhead{$\theta_{0, kep}$}
&\colhead{$B_{m,0,min}$} &\colhead{$B_{m, 0, opt}$}
&\colhead{$B_{\phi, 0, opt}$} &\colhead{$B_{r,\star,L}$}
&\colhead{$t_{circ}$} &\colhead{$t_{fill}$}
&\colhead{$\hat{\rho}_{fill, int}$}\\
\colhead{Type} & &\colhead{($\deg$)} &\colhead{($\deg$)}
&\colhead{(G)} &\colhead{(G)} &\colhead{(G)} &\colhead{(G)}
&\colhead{(years)} &\colhead{(years)} &\colhead{($\textrm{gm
cm}^{-3}$)}} \startdata
O3   & 0.60 &65 &70.7 & $437$ &$893$  & $-63$   & $139$  & $0.53$ & 0.18 &$9.5 \times 10^{-11}$ \\
     &      &75 &78.5 & $540$ &$1100$ & $-46$   & & & &  $6.2 \times 10^{-11}$ \\
O6.5 & 0.60 &65 &70.7 & $101$ &$207$  & $-20$   & $175$  & $0.57$ & 0.12 &$6.0 \times 10^{-12}$ \\
     &      &75 &78.5 &$125$  &$256$  & $-15$   & & & &  $3.9 \times 10^{-12}$ \\
B0   & 0.60 &65 &70.7 & $33$  &$67$   & $-12$   & $131$  & $0.31$ & 0.03 &$5.0 \times 10^{-13}$ \\
     &      &75 &78.5 & $40$  &$83$   & $-9$    & & & &  $3.2 \times 10^{-13}$ \\
B2   & 0.50 &65 &69.6 & $7.0$ &$13$   & $-2.1$  & $25$  & $1.1$  & 0.02 &$8.7 \times 10^{-15}$ \\
     &      &75 &77.8 & $8.6$ &$16$   & $-1.6$  & & & &  $5.4 \times 10^{-15}$ \\
B5   & 0.45 &65 &69.1 & $1.4$ &$2.5$  & $-0.40$ & $3.6$ & $6.7$ & 0.02 &$2.7 \times 10^{-16}$ \\
     &      &75 &77.5 & $1.8$ &$3.1$  & $-0.30$ & & & &  $1.7 \times 10^{-16}$ \\
B9   & 0.40 &65 &68.7 & $0.76$ &$1.3$ & $-0.18$ & $0.5$ & $30.6$ & 0.02 &$7.0 \times 10^{-17}$ \\
     &      &75 &77.3 & $0.94$ &$1.6$ & $-0.13$ & & & &  $4.3 \times 10^{-17}$ \\
\enddata
\tablecomments{We use $\beta = 1.0$ to compute $B_{m, 0, min}$,
$B_{m, 0, opt}$ and $B_{\phi, 0, opt}$. Values of $t_{fill}$ are
computed for $\delta_{fill} = 10^\circ$. Density of the disk at
the point $X_{int}$ at the end of the fill-up stage is
$\hat{\rho}_{int, fill}$}
\end{deluxetable}

\clearpage

\begin{deluxetable}{lccccc}
\tabletypesize{\scriptsize} \tablecaption{Ratios of magnetic force
to centrifugal force and viscous force to magnetic force at the
end of the disk fill-up stage \label{table-stars-Fmag-Fvisc}}
\tablewidth{0pt} \tablecolumns{2} \tablehead{\colhead{Spectral}
&\colhead{$\alpha$} &\colhead{$\theta_{0, lim}$} &
\colhead{$F_{mag}/F_{cen}$}
& \colhead{$F_{visc}/F_{mag}$}\\
\colhead{Type} & &\colhead{($\deg$)} & & } \startdata
O3   & 0.60 &65 &0.005 &0.056\\
     &      &75 &0.005 &0.057\\
O6.5 & 0.60 &65 &0.004 &0.045\\
     &      &75 &0.004 &0.046\\
B0   & 0.60 &65 &0.011 &0.016\\
     &      &75 &0.011 &0.015\\
B2   & 0.50 &65 &0.014 &0.011\\
     &      &75 &0.014 &0.011\\
B5   & 0.45 &65 &0.017 &0.008\\
     &      &75 &0.018 &0.008\\
B9   & 0.40 &65 &0.012 &0.010\\
     &      &75 &0.013 &0.009\\
\enddata
\tablecomments{We take $\nu=0.1$ and $\beta=1.0$.
$F_{mag}/F_{cen}$ is an estimate of the ratio of the magnetic
force to centrifugal force in the disk and is the average value of
$\hat{B}^2/(4 \pi \hat{\rho}_{fill} \hat{v}_\phi^2)$ taken at
$X_{kep}$ and ${X_{lim}}$. $F_{visc}/F_{mag}$ is an estimate of
the ratio of the viscous force to magnetic force and is the
average value of $(4 \pi \hat{\rho}_{fill} \nu
\hat{c}_S^2/\hat{B}^2)$ at the same two points.}
\end{deluxetable}

\clearpage

\begin{deluxetable}{rrrrccccccrrr}
\tabletypesize{\scriptsize} \tablecaption{Critical magnetic field
strengths, extents of optimal disk regions and latitude ranges of
$\Sigma_{0, disk}$ for the B2 star model
\label{table-B2-Bcrit-alpha}} \tablewidth{0pt} \tablecolumns{2}
\tablehead{ \multicolumn{6}{c}{$\theta_{0, lim} = 70^{\circ}$} & &
&\multicolumn{5}{c}{$\alpha = 0.5$} \\
\cline{1-6} \cline{9-13} \\
& &\multicolumn{3}{c}{Optimal model} & & & &
&\multicolumn{3}{c}{Optimal model} &\\
\cline{3-5} \cline{10-12}
\\
\colhead{$\alpha$} &\colhead{$x_{kep}$} &\colhead{$x_{end}$}
&\colhead{$B_{m, 0, opt}$} &\colhead{$B_{\phi, 0, opt}$} &
\colhead{$B_{m,0,min}$} & & & \colhead{$\theta_{0, lim}$} &
\colhead{$\theta_{0, kep}$} &\colhead{$B_{m, 0, opt}$}
&\colhead{$B_{\phi, 0, opt}$}  & \colhead{$B_{m,0,min}$}\\
 & & &\colhead{(G)} &\colhead{(G)} &\colhead{(G)} & &
&\colhead{($\deg$)} & \colhead{($\deg$)} & \colhead{(G)}
&\colhead{(G)} &\colhead{(G)}} \startdata
0.35 & 2.01  & 5.07 & $22.4$  & $-1.4$  & $13.9$ & & &60 &65.4 & $12.0$  & $-2.4$ & $6.6$ \\
0.40 & 1.84  & 4.64 & $18.9$  & $-1.6$  & $11.3$ & & &65 &69.6 & $12.7$  & $-2.1$ & $7.0$ \\
0.45 & 1.70  & 4.29 & $16.2$  & $-1.7$  & $ 9.3$ & & &75 &7.8  & $15.8$  & $-1.5$ & $8.6$ \\
0.50 & 1.59  & 4.00 & $13.9$  & $-1.9$  & $ 7.6$ & & &80 &81.9 & $19.1$  & $-1.3$ & $10.4$ \\
0.55 & 1.49  & 3.75 & $12.0$  & $-2.0$  & $ 6.3$ & & &85 &85.9 & $26.7$  & $-0.9$ & $14.7$ \\
\enddata

\tablecomments{Here $\beta = 1.0$ in all cases.}
\end{deluxetable}

\clearpage

\begin{deluxetable}{rrrrr}
\tabletypesize{\scriptsize} \tablecaption{Variation of surface
magnetic field strengths and latitude ranges of $\Sigma_{0, disk}$
with $x_{lim}$ \label{table-B2-Bm0-Bphi0-xlim}} \tablewidth{0pt}
\tablecolumns{2} \tablehead{ \colhead{$x_{lim}$} &
\colhead{$x_{end}$} &\colhead{$B_{m, 0, lim}$}
&\colhead{$B_{\phi, 0, lim}$} &\colhead{$\theta_{kep}$} \\
& &\colhead{(G)} &\colhead{(G)}  &\colhead{($\deg$)}} \startdata
1.62   & $1.72$ & $8.1$   & $-1.75$ & 70.4  \\
2.02   & $4.00$ & $13.9$  & $-1.85$ & 73.7  \\
3.18   & $4.00$ & $32.8$  & $-2.26$ & 77.3  \\
4.77   & $4.00$ & $61.4$  & $-2.86$ & 79.0  \\
\enddata
\tablecomments{Values shown are for models of the B2 star with
$\alpha = 0.5$, $\theta_{0,lim}=70^\circ$ and $\beta = 1.0$. In
all these models, $x_{kep} = 1.59$ and $x_{esc} = 2.00$.}
\end{deluxetable}

\end{document}